\documentclass[11pt]{article}
\pdfoutput=1
\usepackage{jheppub} 

\usepackage[T1]{fontenc} 
\usepackage{braket}
\usepackage{latexsym}
\usepackage{wrapfig}
\usepackage{tabularx}
\usepackage{comment}
\newcommand{\coloneqq}{:=}
\newcommand{\der}{\partial}
\newcommand{\no}{\nonumber}

\newcommand{\im}{\mathrm{Im}}

\newcommand{\Mpl}{M_{\rm pl}}
\newcommand{\Ms}{M_{\rm s}}
\newcommand{\Mstar}{M_*}

\newcommand{\mth}{m_\text{th}}
\newcommand{\scat}{\mathcal{M}}

\newcommand{\J}{\mathcal{J}}

\newcommand{\uv}{\text{$s\sim\Ms^2$}}
\newcommand{\uvir}{\text{$\delta<y<1$}}

\newcommand{\nullc}{\mathcal{N}}
\newcommand{\llangle}{\langle}
\newcommand{\rrangle}{\rangle}

\title{
Finite energy sum rules for gravitational Regge amplitudes
}

\author[a]{Toshifumi Noumi}
\author[a,b]{and Junsei Tokuda}

\affiliation[a]{Department of Physics, Kobe University,
 Kobe 657-8501, Japan}
\affiliation[b]{Center for Theoretical Physics of the Universe, Institute for Basic Science (IBS),
  Daejeon, 34126, Korea}

\emailAdd{tnoumi@phys.sci.kobe-u.ac.jp}
\emailAdd{jtokuda@ibs.re.kr}

\abstract{
We develop a framework to derive consistency constraints on gravitational Regge amplitudes based on the finite energy sum rules (FESRs), which directly connect gravitational Regge amplitudes at a finite ultraviolet scale with infrared physics without suffering from super-Planckian physics. For illustration, we consider four-point scattering of an identical massless scalar coupled to gravity. First, we derive multiple FESRs without relying on the $s\text{-}t\text{-}u$ permutation invariance. We then make use of FESRs, crossing symmetry, and other principles such as unitarity, to derive bounds on the Regge parameters. The bounds result in infrared finite gravitational positivity bounds in four spacetime dimensions.

}

\begin{document}
{\baselineskip0pt
\rightline{\baselineskip16pt\rm\vbox to-20pt{
           \hbox{KOBE-COSMO-22-20}
           \hbox{CTPU-PTC-22-29}
\vss}}%
}

\maketitle
\flushbottom

\section{Introduction}

How can we explore the nature of quantum gravity by experiments? Ideally, we would like to probe a phenomenon whose energy scale is as high as the Planck scale $\Mpl$, but it is too high to access directly.
However, recent progress in the Swampland Program~\cite{Vafa:2005ui} points an interesting possibility that the nature of quantum gravity may show up in particle physics and cosmology through hidden quantum gravity constraints on low-energy effective field theories (EFTs). Such swampland conditions have been proposed and studied with various degrees of rigours and motivations. See~\cite{Brennan:2017rbf, Palti:2019pca, vanBeest:2021lhn} for review articles.

Positivity bounds on low-energy scattering amplitudes offer a useful tool to derive such ultraviolet (UV) constraints on low-energy EFTs~\cite{Pham:1985cr,Adams:2006sv}. They are based on fundamental properties of the S-matrix such as unitarity, analyticity and the mild high-energy behavior in the Regge limit, the former two of which are summarized into the twice-subtracted dispersion relation in particular. Those UV properties of scattering amplitudes provide various inequalities among Wilson coefficients as a necessary condition for a low-energy EFT to have a standard UV completion.

While positivity bounds are originally formulated in nongravitational theories and well established in gapped theories, their generalization to gravity theories is crucial for the Swampland Program.
For example, the bounds on tree-level light-by-light scattering in the graviton-photon EFT imply that macroscopic extremal black holes satisfy the Weak Gravity Conjecture bound~\cite{Arkani-Hamed:2006emk}. See, {\it e.g.},~\cite{Hamada:2018dde,Bellazzini:2019xts,Arkani-Hamed:2021ajd,Noumi:2022ybv}. 
Furthermore, if the bounds are applicable to the matter-matter scattering amplitudes at loop level in four spacetime dimentions up to $\mathcal{O}(\Mpl^{-2})$, one may derive constraints on the spectrum and interactions of light particles well below the Planck scale~\cite{Cheung:2014ega,Andriolo:2018lvp,Chen:2019qvr,Alberte:2020bdz,Aoki:2021ckh,Noumi:2021uuv,Noumi:2022ybv,Noumi:2022zht}.

This motivation led to various attempts to formulate positivity bounds on gravitational theories.
An obvious obstruction is the absence of proof of twice-subtracted dispersion relation. For recent discussions on this point, see~\cite{Haring:2022cyf}. The twice-subtracted dispersion relation for four-point scattering amplitudes $\scat(s,t)$ with negative momentum transfer $t<0$ is often assumed in the literature and satisfied in known examples, {\it i.e.,} tree-level string amplitudes. We simply use this property as a postulate in this work. Interestingly, it is still nontrivial to derive positivity bounds due to the presence of graviton $t$-channel pole.

To avoid the $t$-channel pole issue, \cite{Caron-Huot:2021rmr} proposed to work in finite impact parameter $b\sim M^{-1}$ to make the contributions of graviton finite. This method works in higher spacetime dimensions $D>4$ and it was found that the coefficient $c_2(0)$ of the $s^2$ term in the IR amplitude (see eq.~\eqref{IRexp} for its definition) can be negative but the modulus of negativity is bounded from below by the scale of new physics $M$. It also proved an expected scaling that Wilson coefficients of higher-derivative corrections are suppressed by the scale of new physics $M$. 
Intuitively, this would simply state that the length scales of interactions between light particles mediated by heavy physics cannot exceed the Compton wavelength of heavy mediators and the dimensionless light-heavy coupling strength cannot exceed $\mathcal{O}(1)$ due to the unitarity bounds. 

However, their method suffers from infrared (IR) divergences in $D=4$ when moving from momentum space to the impact parameter space. As a result, the bounds have logarithmic dependence on the IR cutoff~\cite{Caron-Huot:2021rmr,Caron-Huot:2022ugt}. This IR divergence comes from graviton $1/t$ pole. Although this is a physical singularity, the IR divergence in the bound obtained via their method may be simply a technical problem: physically, it would be natural to expect the existence of the bounds in $D=4$ which are similar to those derived in higher dimensions. It is desirable to derive bounds in $D=4$, especially for phenomenological application.

Besides the finite-$t$ program mentioned above, another methodology based on the Regge behavior has been proposed in~\cite{Hamada:2018dde,Tokuda:2020mlf} and discussed in~\cite{Herrero-Valea:2020wxz, Alberte:2021dnj,Herrero-Valea:2022lfd}. In this method, the graviton $t$-channel pole in the dispersive sum rule for $c_2(0)$ is canceled with the dispersive integral of the Regge amplitude $\scat(s,t)\sim f(t)s^{2+\alpha't+\alpha''t^2/2+\cdots}$ up to some finite $\mathcal{O}(t^0)$ residuals: for details, see sec.~\ref{sec:review}. The Regge behavior is realized by the tower of higher-spin states whose onset is $\Ms$, which corresponds to the string scale in the perturbative string.
It was found in \cite{Tokuda:2020mlf} that such finite terms include $\mathcal{O}(\Mpl^{-2}M^{-2})$ negative terms which are obstructions for ruling out a tiny amount of negativity, 
that is consistent with \cite{Alberte:2020jsk}. Here, the scale $M$ is determined once the details of the Regge behavior are given.

An advantage of this approach is that it does not suffer from the IR divergence even in $D=4$ case up to $\mathcal{O}(\Mpl^{-2})$. On the other hand, there is also a disadvantage that the scale $M$ is unknown unless the details of the Regge behavior at UV are given: ultimately the bound becomes meaningless if the scale $M$ can be arbitrarily small although it seems plausible to assume $M\sim\Ms$ as long as the Regge behavior is governed by the exchange of heavy higher-spin states. This issue becomes more subtle if loops of light particles are included:\footnote{Note that loop corrections from higher-derivative vertices may not affect the tree-level results so much~\cite{Bellazzini:2021oaj}.}  see {\it e.g.,} \cite{Alberte:2021dnj} for recent discussions for photon-graviton scatterings. 
Apart from the purely theoretical study, the interesting possibility of exploring the properties of quantum gravity such as the scale $M$ via experimental search of dark sector physics has been pointed out quantitatively in~\cite{Noumi:2022zht}.

In this paper, we derive constraints on the scale $M$ to further develop the latter method based on the Regge behavior as a tool to provide interesting bounds in $D=4$ spacetime dimensions. As a proof of concept, we focus on the two-to-two scattering of an identical massless scalar $\phi$ coupled to gravity. To determine the scale $M$, we need to know the details of the Regge behavior. 
It has been known that the so-called finite-energy sum rules (FESRs) are useful to constrain the Regge behavior from low-energy data in the context of physics of strong interactions~\cite{Igi:1962zz,Logunov:1967dy,Igi:1967zza,Gatto:1967zza,Dolen:1967zz,Dolen:1967jr}. This is because, FESRs express the Regge parameters in terms of the dispersive integral below the Reggeization scales and the latter can be fixed by low-energy measurements alone. Historically, this idea was proposed by Igi in \cite{Igi:1962zz} and the more detailed analysis in~\cite{Dolen:1967zz,Dolen:1967jr} lead to the finding of so-called Dolen-Horn-Schmidt (DHS) duality, the novel feature which is not explained by the ordinary Feynman diagrams. Soon after this finding, the works \cite{Ademollo:1967zz,Ademollo:1968cno} based on the duality and FESRs gave a tantalizing hint for the existence of Veneziano amplitude~\cite{Veneziano:1968yb}, which is now known as the tree-level string amplitude.

We rekindle this classic idea of FESRs to get more information of the Reggeization of graviton exchange which is realized by higher-spin towers. 
Although we do not have any experimental data for higher-spin particles above $\Ms$, we can use the null constraints derived from crossing symmetry as an input. We then derive constraints on the Regge parameters and the scale $M$, leading to the manifestly IR finite gravitational positivity bounds in $D=4$ spacetime dimensions. This is the key idea of this work. We find the scaling $M\gtrsim \Ms$ when ignoring loops of light particles. We also discuss how this result changes once such loop contributions are taken into account.

This paper is organized as follows: in sec.~\ref{sec:review}, we briefly explain the original derivation of gravitational positivity bounds based on the Regge behavior, which is developed in~\cite{Tokuda:2020mlf}. We also provide some technical comments. In sec.~\ref{sec:null}, we review the null constraints and then discuss the sign of their IR part. The resultant bounds are useful for our purpose. In sec.~\ref{sec:FESR}, we derive FESRs for the Regge parameters associated with graviton exchange. We use them to derive constraints on the Regge parameters and the scale $M$ in sec.~\ref{sec:FESRbound}. We provide several remarks on our formalism in sec.~\ref{sec:remark}.
We then conclude in sec.~\ref{sec:concl}.
Some technical details are collected in appendices.

\section{Gravitational positivity bounds} \label{sec:review}

We consider a low-energy EFT of a massless scalar $\phi$ and the graviton, and its UV completion. We denote the four-point scattering amplitude of $\phi$ in the UV complete description by $\scat(s,t)$. In the present analysis, we ignore loops of light particles except in sec.~\ref{sec:loop}. Hence, we work up to $\mathcal{O}(\Mpl^{-2})$. While we do not specify details of the UV complete theory, we postulate several UV properties of the scattering amplitude $\scat(s,t)$ and discuss their implications for the UV constraints on the IR physics.

As we mentioned in the introduction, general UV properties of scattering amplitudes in quantum gravity are still unknown. We thus {\it postulate} that $\scat(s,t)$ is unitary and analytic,
and it has a mild UV behavior $\lim_{|s|\to\infty}|\scat(s,t<0)/s^2|=0$.
Then, the location of massive singularities of $\scat$ on the complex $s$-plane can be identified with the mass scale of new physics beyond the scalar-graviton EFT. We define this scale by $\Ms$, which turns out to be analogous to the string scale in our tree-level working assumption. 
Above the scale $\Ms$, unknown heavy states such as tower of higher-spin states appear and EFT description breaks down. Note that we can add any particles whose spins are less than two to our EFT without affecting our result, as long as they do not invalidate the tree-level approximation in the EFT. This is because we can simply subtract poles associated with particles with spin lower than two without affecting the dispersion relation derived below.

 Now we define the low-energy expansion coefficients $\{c_n(t)\}$ as 
\begin{align}
\scat(s,t)-(s,t,u\text{-channel\,\,poles})=\sum_{n=0}^\infty\frac{c_{2n}(t)}{(2n)!}\left(\frac{s-u}{2}\right)^{2n}=\sum_{n=0}^\infty\frac{c_{2n}(t)}{(2n)!}\left(s+\frac{t}{2}\right)^{2n}\,.
\label{IRexp}
\end{align}
Note that $(su/\Mpl^2t)\in(t\text{-channel\,\,poles})$ expresses the graviton $t$-channel pole. The properties mentioned above give rise to the following sum rule for $c_2(t)$:
\begin{align}
    c_2(t)
    =
    \frac{4}{\pi}\int^\infty_{\Ms^2}\mathrm{d}s\,\frac{\im\,\scat(s,t)}{\left(s+(t/2)\right)^3}
    + \frac{2}{\Mpl^2t}
    \,.\label{eq:disp1}
\end{align}
To ensure the regularity of the left-hand side (LHS) in the forward limit, $\im\,\scat$ must grow at least as fast as $s^2$ in the limit $t\to-0$.\footnote{This will be true provided that $\im\,\scat(s,t)$ is regular for $(s,t)$ satisfying $s\geq\Ms^2$ and $-\delta<t\leq0$, where $\delta$ is some infinitesimal positive constant. This condition will be satisfied when we ignore loop corrections from massless particles such as gravitons.}
This naturally motivates the Regge behavior\footnote{In general we can also add sub-leading terms to \eqref{eq:regge1}. We briefly discuss how the presence of such terms may modify our analysis in sec.~\ref{sec:subleading}.}
\begin{align}
    \im\,\scat(s,t)
    \simeq \im\,\scat_\text{R}(s,t)=f(t)\left(\frac{s+(t/2)}{\Ms^2/\epsilon+(t/2)}\right)^{\alpha(t)}
    \quad (\text{for} \quad s\geq\Mstar^2>\Ms^2)\,,
\label{eq:regge1}
\end{align}
where $\alpha(t)=2+\alpha't+\alpha''t^2/2+\cdots$ with $\alpha'>0$ and $\Ms$ denotes the lightest mass of heavy states including tower of higher-spin states which realize the Regge behavior. We also introduce a tiny positive constant $\epsilon\leq(\Ms^2/\Mstar^2)<1$ in \eqref{eq:regge1} just for a notational convenience. 
Then, we perform the integral on the right-hand side (RHS) of eq.~\eqref{eq:disp1} from $s=\Ms^2/\epsilon$ to $s=\infty$ by using \eqref{eq:regge1} and take the $t\to-0$ limit to get
\begin{align}
	c_2(0)
	\simeq
	\frac{4}{\pi}\int^{\Ms^2/\epsilon}_{\Ms^2}\mathrm{d}s\,\frac{\im\,\scat(s,0)}{s^3}
	+F_0\,,
    \qquad 
    F_0
	&\coloneqq
	-\frac{1}{\Mpl^2}
		\left[
			\frac{2f'}{f}
            -\frac{2\epsilon}{\Ms^2}
			-\frac{\alpha''}{\alpha'}
		\right]
    \,. \label{eq:disp2}
\end{align}
Here, $f\coloneqq f(0)$ and $f'\coloneqq \der_t f(t)|_{t=0}$. 
Note that the condition $f\epsilon^2=\pi\Ms^4\alpha'/(2\Mpl^{2})$ is required by the absence of $t^{-1}$-singularity in \eqref{eq:disp2}.
The condition $f'/f\geq0$, which is imposed by unitarity, explains why $c_2(0)$ can be negative in general. Let us denote the scaling of the $t$-dependence of $f(t)$ and $\alpha(t)$ by 
the scale $M$ as $f'/f\,,\,
	-\alpha''/\alpha'
	\lesssim
	\mathcal{O}(M^{-2})$. 
 Then, \eqref{eq:disp2} reads 
 \begin{align}
    c_2(0)
    > F_0 > -\mathcal{O}(\Mpl^{-2}M^{-2})
    \,.
    \label{eq:disp3}
\end{align}
Our goal is to prove the scaling,
\begin{align}
    M\gtrsim \Ms
\,,\label{eq:scaling1}
\end{align} 
including a numerical factor, when ignoring the loops of light particles, by providing a lower bound on $F_0$. We also discuss the case where such loops are taken into account in sec.~\ref{sec:loop}.

\paragraph{A comment on $\epsilon$-dependence.}
A brief comment on the $\epsilon$-dependence of the sum rule will be useful. A precise value of $F_0$ depends on the $\epsilon$, while the RHS of \eqref{eq:disp2} can be shown to be independent of $\epsilon$~\cite{Tokuda:2020mlf}.\footnote{For this purpose, it is more convenient to parameterize $\im\,\scat_\text{R}$ in an $\epsilon$-independent manner unlike \eqref{eq:regge1}.}  This means that, combined with the positivity condition $\im\,\scat(s,0)>0$, 
$F_0$ is a monotonically decreasing function of $\epsilon$. 
Hence, we can take $\epsilon$ arbitrarily as long as \eqref{eq:regge1} is a good approximation. To optimize \eqref{eq:disp3}, we should take $\epsilon$ as large as possible. We thus take $\epsilon=\Ms^2/\Mstar^2<1$, the possible largest choice of $\epsilon$. We in general expect that the amplitude is Reggeized at energy scales above which many higher-spin states are excited. In known Regge amplitudes, such as string amplitudes, we have $\Ms^2/\Mstar^2=\mathcal{O}(0.01)$. We will then take $\epsilon=0.01$ as a benchmark point later. We stress however that the $\epsilon$-dependence of our final bounds on $F_0$ is weak so that our main conclusion is insensitive to the choice of $\epsilon$.

\section{Null constraints and low-energy pieces}\label{sec:null}
We first introduce null constraints known in the literature in sec.~\ref{sec:nullreview}. We then discuss the IR part of the null constraints in sec.~\ref{sec:appnull}. The resultant bounds turn out to be useful to constrain the Regge parameters such as $\alpha(t)$ in the succeeding sections.

\subsection{Null constraints}\label{sec:nullreview}
Crossing symmetries imply the $s\text{-}t\text{-}u$ permutation invariance of $\scat(s,t)$. This imposes nontrivial relations among different low-energy coefficients. As a result, the relations among sum rules for different coefficients are also imposed, leading to the so-called null constraints. The null constraints have been used to tighten positivity bounds in nongravitational setup~\cite{Arkani-Hamed:2020blm,Bellazzini:2020cot,Caron-Huot:2020cmc,Tolley:2020gtv,Sinha:2020win}. Also, the null constraints are used in the impact-parameter method to derive positivity bounds in the presence of gravity in higher dimensional spacetime proposed by \cite{Caron-Huot:2021rmr}. The sum rules for coefficients of $s^nt^m$ with $n\geq4$ are unaffected by the graviton $t$-channel pole. Hence, higher-order null constraints which are associated with crossing relations imposed on such coefficients can be derived just as in nongravitational theories. Useful sum rules which generate null constraints systematically are~\cite{Caron-Huot:2020cmc,Caron-Huot:2021rmr} 
\begin{align}
	&\llangle X_\ell(t;s,J)\rrangle
	=
	0\qquad
	(\ell=4,6,\cdots)\,,\label{nullgen1}
	\\
	&X_\ell(t;s,J)
	\coloneqq 
	\frac{2s+t}{t(s+t)}\frac{P_J(1+\frac{2t}{s})}{(ts(s+t))^{\ell/2}}
	\no\\
	&\qquad\qquad\qquad
	-\mathop{\text{Res}}_{x=0}
		\left[
			\frac{(2s+x)(s-x)(s+2x)}{x(t-x)(s+x)(s-t)(s+t+x)}
			\frac{P_J(1+\frac{2x}{s})}{(xs(x+s))^{\ell/2}}
		\right]
	\,.\label{nullgen_def}
\end{align}
Here, we define the average $\llangle(\cdots)\rrangle$ with respect to the nonnegative spectral density $\rho_J(s)$ as
\begin{align}
	&\llangle(\cdots)\rrangle
	\coloneqq
	\int^{\infty}_{\Ms^2}\,\frac{\mathrm{d}s}{s}
		\sum_{\text{even}J\geq0}n_J\rho_J(s) \,(\cdots)
	\,,\label{average}\\
	&\im\,\scat(s,t)
	=
	\sum_{\text{even}J\geq0}
	n_J\rho_J(s)P_J
	\left(
		1+\frac{2t}{s}
	\right)
	\,,\quad
	 n_J
	 = (16\pi)(2J+1)\,.\label{partialwave}
\end{align}
We define the null constraints derived from the sum rules $\llangle\der_t^{k-3}X_4(t;s,J)|_{t=0}\rrangle=0$ $(k=3,4,5,\cdots)$ as $\llangle\nullc_{k}(\J^2)\,s^{-4-k}\rrangle=0$, where $\J^2\coloneqq J(J+1)$. 
$\nullc_k(\J^2)$ is a polynomial of $\J^2$ of degree $k$ and we impose the normalization condition such that $\nullc_k(\J^2)\to2(\J^2)^k$ in the large $J$ limit: for example, $\nullc_3(\J^2)=\J^2(\J^2-6)(2\J^2-49)$. Explicit expressions for $\nullc_k$ with $k=4,5,6$ are shown in app.~\ref{sec:nullexp} as concrete examples. 

$\nullc_k(\J^2)$ is positive definite at large $J$, while it can be negative at low $J$. For example, $\nullc_3(\J^2)=0$ for $J=0,2$ and $\nullc_3(\J^2)<0$ for $J=4$, while $\nullc_3(\J^2)>0$ for $J=6,8,\cdots$. Then the $\nullc_3$-constraint provides a non-trivial balance between the low-spin contribution and the higher-spin contributions.  This feature holds in general for $\nullc_k$-constraints with $k=3,4,5,\cdots$. 

\subsection{IR part of null constraints } \label{sec:appnull}
We can write the $\nullc_k$-constraints in terms of $\im\,\scat$ as 
\begin{align}
    \llangle s^{-4-k}\nullc_k(\J^2)\rrangle
    =
    \int^\infty_{\Ms^2}\mathrm{d}s\,s^{-5-k}\sum_{n=1}^kc_{k,n}(s\der_t)^n\im\,\scat(s,t)|_{t=0}
    = 0\,,
    \label{nullcexp1}
\end{align}
by using the fact that $(s\der_t)^n P_J(1+\frac{2t}{s})$ is a polynomial in $J$ of degree $2n$ with $n\leq J$. Here, $c_{k,n}$'s are dimensionless constants and the normalization of $\nullc_k$ concludes $c_{k,k}>0$: {\it e.g.}, we have $(c_{3,3},c_{3,2},c_{3,1})=(12,-90,180)$.
We can evaluate the high-energy pieces of the integral on the RHS by using the Regge amplitude $\im\,\scat_\text{R}$. In particular, let us consider the integration from $s=\Ms^2/\epsilon$ to $s=\infty$. For sufficiently tiny $\epsilon$, the term with $n=k$ becomes dominant because $s^{-5-k}(s\der_t)^n\im\scat_\text{R}|_{t=0}\sim s^{-3-(k-n)}[\ln(s)]^n$. As a result, the contributions to the RHS of \eqref{nullcexp1} from high energy regions $s>\Ms^2/\epsilon$ become positive definite because of the condition $c_{k,k}>0$ and the positivity $\der_t^k\im\scat_\text{R}|_{t=0}>0$ implied by unitarity. This means that the following inequalities hold for sufficiently tiny $\epsilon\ll1$,
 \begin{align}
     \langle s^{-4-k}\nullc_k(\J^2)\rangle_\uv
     =
     - \int^\infty_{\Ms^2/\epsilon}\mathrm{d}s\,s^{-5-k}\sum_{n=1}^kc_{k,n}(s\der_t)^n\im\,\scat(s,t)|_{t=0}
     \leq0
    \,,\label{nullapp}
 \end{align}
where we introduced the following notation:
\begin{align}
	\llangle(\cdots)\rrangle_\uv
	&\coloneqq 
	\int^{\Ms^2/\epsilon}_{\Ms^2}\,\frac{\mathrm{d}s}{s}
	\sum_{\text{even}J\geq0}n_J\rho_J(s) \,(\cdots)
	\label{midavs}\\
	&=
	\int^{1}_{\epsilon}\,\frac{\mathrm{d}y}{y}
	\sum_{\text{even}J\geq0}n_J\rho_J(y\Ms^2/\epsilon) \,(\cdots)
	\,,\quad y\coloneqq \frac{s}{\Ms^2/\epsilon}
	\,.\label{midavy}
\end{align}
When terms $(\cdots)$ inside the angle bracket $\langle\rangle$ is written in terms of $y$, the average should be understood as \eqref{midavy}. 

Physically, the relation \eqref{nullapp} may follow from the simple intuition: null constraints \eqref{nullcexp1} require that the contributions from the IR part of the integral are exactly canceled with those from UV part. These contributions must have opposite sign with each other. Large-$J$ states which positively contribute to $\nullc_k(\J^2)$ are effectively excited only at UV, while low-$J$ states negatively contribute to $\nullc_k$. As a result, the UV (IR) part of the integral tend to be positive (negative). As explained in sec.~\ref{sec:review}, we choose $\epsilon=\Ms^2/\Mstar^2$. We suppose that \eqref{nullapp} is also valid for this choice of $\epsilon$.

We emphasize that the derivation of \eqref{nullapp} is merely based on the simple high-energy behavior of the amplitude. It is not necessary to use any special properties of gravitational or Regge amplitudes for motivating \eqref{nullapp}: we also expect the validity \eqref{nullapp} in the context without gravity. For instance, we can confirm that \eqref{nullapp} is satisfied by a scalar box amplitude $\scat_\text{box}$ because we have $s^{-5-k}(s\der_t)^n\im\,\scat_\text{box}\sim s^{-6-(k-n)}$ at high energies. We can also show the validity of \eqref{nullapp} for string amplitudes: see app.~\ref{sec:checkappnull} for more details on these points.

\section{Finite energy sum rules (FESRs)}\label{sec:FESR}
Now we derive the multiple finite energy sum rules (FESRs). In sec.~\ref{sec:FESRderiv}, we derive FESRs for $f'$ and $\alpha''$. Several useful relations are derived in sec.~\ref{sec:usefulbound}. Such relations and the $\nullc_k$-constraint \eqref{nullapp} are then used to derive the bounds on $f'$ and $\alpha''$ in the next section. We note that the derivation of our FESRs can be straightforwardly extended to the processes which are not invariant under the $s\text{-}t\text{-}u$ permutation. 

\subsection{Finite energy sum rules}\label{sec:FESRderiv}
\begin{figure}[tbp]
 \centering
  \includegraphics[width=.7\textwidth, trim=150 220 120 140,clip]{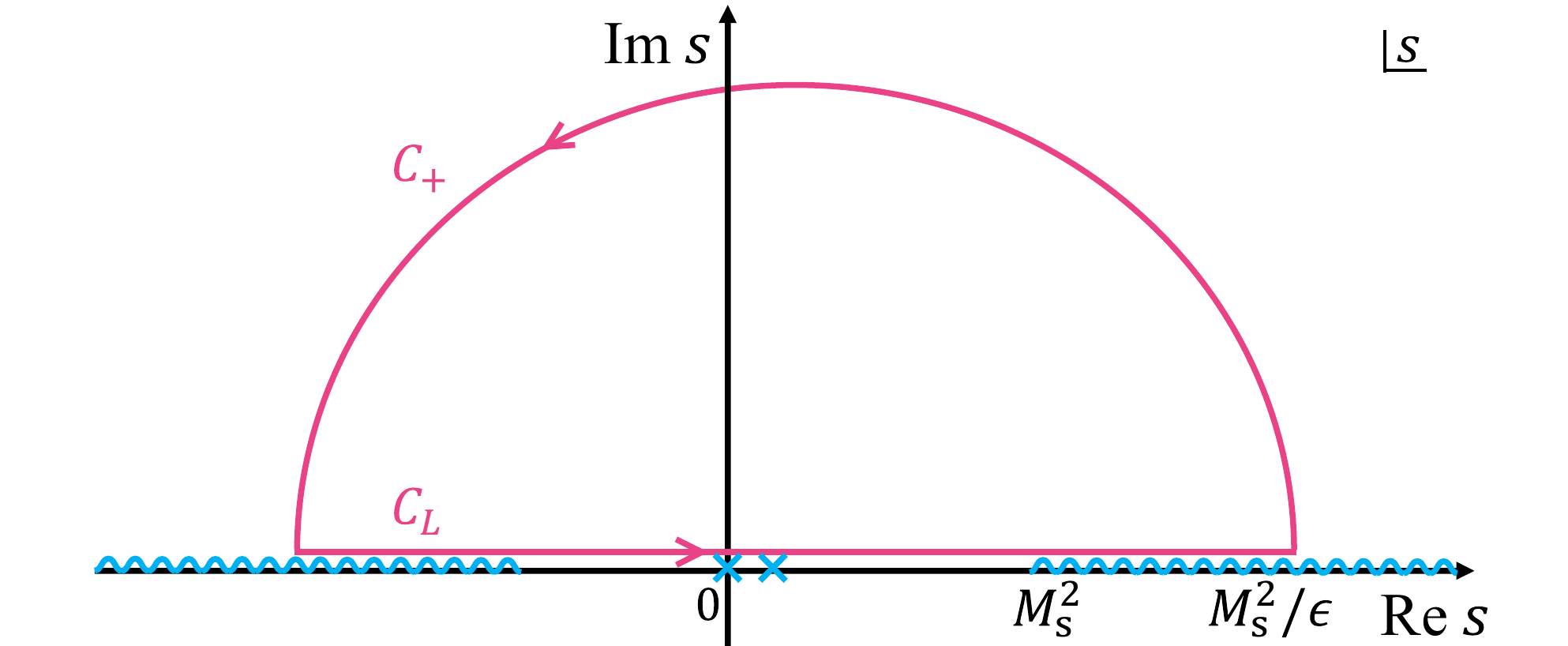}
 \caption{The integration contour on the complex $s$-plane which is considered in \eqref{start1} to derive the FESRs. A semi-circle on the upper half plane $\mathcal{C}_+$ is centered at $s$-$u$ crossing symmetric point $s=-t/2$ and the radius is $\Ms^2/\epsilon+(t/2)$. The wavy lines and the $``\times"$ represent the branch cuts and poles of $s,u$-channel graviton exchange, respectively. We do not assume that $\scat$ is analytic outside the contour.
}
 \label{fig:FESRcontour} 
\end{figure}

Let us consider the following complex integrals of $\scat$ and their implications:
\begin{align}
\oint_{\mathcal{C}_++\mathcal{C}_L}\frac{\mathrm{d}s}{2\pi i}\,\scat(s,t)(s+(t/2))^{2n+1}=0\qquad (n=0,1,2,\cdots)\,,\label{start1}
\end{align}
where the integration contour $\mathcal{C}_++\mathcal{C}_L$ is shown in fig.~\ref{fig:FESRcontour}. The contour $\mathcal{C}_+$ is a semi-circle on the upper half plane $\mathcal{C}_+$  centered at the $s$-$u$ crossing symmetric point $s=-t/2$ and its radius is $\Ms^2/\epsilon+(t/2)$. We emphasize that the analyticity of $\scat$ outside the contour $\mathcal{C}_++\mathcal{C}_L$ is not imposed. Using the $s\text{-}u$ crossing symmetry, we can recast \eqref{start1} into the form,
\begin{align}
    -\int_{\mathcal{C}_+}\frac{\mathrm{d}s}{2\pi i}\,\scat(s,t)(s+(t/2))^{2n+1}
    =
    \frac{1}{\pi}
    \int^{\Ms^2/\epsilon}_{-t/2}\mathrm{d}s\,\im\,\scat(s,t)(s+(t/2))^{2n+1}\label{FESRorigin}
\end{align}
for $n=0,1,2,\cdots$. 
We assume that we can evaluate the LHS by using the Regge amplitude $\scat_\text{R}$ in a good approximation:\footnote{The arc $\mathcal{C}_+$ contains the region close to the real $s$-axis. In this region, the amplitude may not be well approximated by $\scat_\text{R}$ particulartly when poles are present in the vicinity of the real axis. This error will be reduced by choosing the location of the arc $\mathcal{C}_+$ far away from the nearest pole as possible. Indeed, the FESRs with this choice of $\mathcal{C}_+$ are well satisfied by the amplitudes consisting of infinite number of poles~\cite{Ademollo:1968cno,Veneziano:1968yb}. As an illustrative example, we confirm that the FESRs are indeed well satisfied by the string amplitude in app.~\ref{sec:FESRconsitency}.}
\begin{subequations}
\label{spin-2domi}
\begin{align}
    & -\int_{\mathcal{C}_+}\frac{\mathrm{d}s}{2\pi i}\,\scat(s,t)(s+(t/2))^{2n+1}
    \simeq
     -\int_{\mathcal{C}_+}\frac{\mathrm{d}s}{2\pi i}\,\scat_\text{R}(s,t)(s+(t/2))^{2n+1}
     \,,\\
     &\scat_\text{R}(s,t)=\frac{-f(t)\left(e^{-i\pi\alpha(t)}+1\right)}{\sin\left(\pi\alpha(t)\right)}\,\left(\frac{s+(t/2)}{\Ms^2/\epsilon+(t/2)}\right)^{\alpha(t)}\,.
\end{align}    
\end{subequations}
Here we wrote the Regge amplitude in a manifestly $s$-$u$ symmetric form.
We include only the leading-order terms on the RHS of \eqref{spin-2domi}: we assume the dominance of the Regge pole with the spin-2 Regge intercept. Under this leading-order approximation, eq.~\eqref{FESRorigin} reduces to the so-called finite-energy sum rules (FESRs) of the form,
\begin{align}
	S_{2n+1}(t)
	=
	\frac{f(t)}{\alpha(t)+2n+2}
    \qquad 
	(n=0,1,2,\cdots)
	\,,\label{FESR1}
\end{align}
where $S_{2n+1}(t)$ is defined by 
\begin{align}
	&S_{2n+1}(t)
	\coloneqq 
		\left[
			\Ms^2/\epsilon+(t/2)
		\right]^{-2n-2}
	\int^{\Ms^2/\epsilon}_{\Ms^2}\,\mathrm{d}s\,
		\left(
			s+(t/2)
		\right)^{2n+1} 
	\im\,\scat(s,t)
	+ p_n(t)
	\,,\label{eq:Sdef}\\
	&p_n(t)
	\coloneqq
		\left(
			\frac{(t/2)}{\Ms^2/\epsilon+(t/2)}
		\right)^{2n+1}
	\frac{\pi\,\text{Res}_{s=0}\scat(s,t)}{(\Ms^2/\epsilon+(t/2))}
	\,,
        \quad
        \text{Res}_{s=0}\scat(s,t)\sim \frac{-t^2}{\Mpl^2}\,.
\end{align}
The second term $p_n(t)$ expresses contributions from graviton $s,u$-channel poles. Note that there can exist additional contributions to $p_n(t)$ from exchange of light particles other than graviton in generic setups, but such poles can be subtracted without affecting the properties of $\scat(s,t)$ used so far. Hence, the discussions below can be extended to such circumstances.
Note that the definition of $S_{2n+1}$ itself can be extended to negative $n$, while the FESRs \eqref{FESR1} are applicable only for $n=0,1,2,\cdots$. 
FESRs \eqref{FESR1} are useful to constrain the Regge amplitude. For instance, we can derive sum rules for the Regge trajectory from \eqref{FESR1} as 
\begin{align}
\alpha(t)=\frac{(2m+2)S_{2m+1}(t)-(2n+2)S_{2n+1}(t)}{S_{2n+1}(t)-S_{2m+1}(t)}\qquad (n,m=0,1,\cdots)\,.\label{trajFESR1}
\end{align}
Substituting these expressions back into \eqref{FESR1}, we can also obtain the sum rules for $f(t)$. In this way, we can derive sum rules for the Regge trajectory and the Regge residue which are useful for constraining the Regge amplitude. The FESRs have been useful in the context of the physics of strong interactions: see {\it e.g.,} \cite{Igi:1962zz,Logunov:1967dy,Igi:1967zza,Gatto:1967zza,Dolen:1967zz,Dolen:1967jr,Ademollo:1967zz,Ademollo:1968cno}. The expressions \eqref{FESR1} relate the gravitational Regge parameters to the low-energy data below the Reggeization scale. Although we used the $s\text{-}u$ permutation invariance to derive our FESRs \eqref{FESR1} since the four-point amplitude of an identical scalar is considered here, we emphasize that it is straightforward to obtain the FESRs even for non-crossing symmetric processes by just treating the contributions from the left cut and the right cut separately when evaluating \eqref{start1}. 

For our purpose, it is more convenient to directly derive sum rules for $f'$ and $\alpha''$ to put bounds on $F_0$. The FESRs for $S'_{2n+1}(0)\coloneqq \der_t S_{2n+1}(t)|_{t=0}$ read
\begin{align}
S'_{2n+1}(0)=\frac{1}{4(n+2)^2}\left[(2n+4)f'-\alpha'f\right]\,.\label{dotFESR1}
\end{align}
The relations \eqref{dotFESR1} allow us to express $f'$ in terms of $S'_{2n+1}(0)$ as
\begin{align}
&f'=\frac{2}{n-m}\left[(n+2)^2S'_{2n+1}(0)-(m+2)^2S'_{2m+1}(0)\right]\label{alphaSR1a}
\end{align}
for two different nonnegative integers $m$ and $n$.
For example, for $(m,n)=(0,1)$, we have\footnote{Note that the term $p_n(t)$ in \eqref{eq:Sdef} does not appear in \eqref{alphaSR2a} because $\der_t^mp_n(t)|_{t=0}=0$ for $m\leq 2n+2$. For the same reasoning, $p_n(t)$ is irrelevant in the analysis below. }
\begin{align}
	&f'
	= \frac{\epsilon}{\Ms^2}
	\left\langle\,
				y\left(
					-36y^3+27y^2+8y-4
				\right)
			+y(18y^2-8)\J^2
	\right\rangle_\uv
	\,.\label{alphaSR2a}
\end{align}
For $(m,n)=(0,2)$, we have
\begin{align}
	&f'
	= \frac{\epsilon}{\Ms^2}
	\left\langle\,
				y\left(
					-48y^5+40y^4+4y-2
				\right)
			+y(16y^4-4)\J^2
	\right\rangle_\uv
	\,.\label{f'FESR2}
\end{align}
It is also straightforward to derive sum rules for $\alpha'$ as
\begin{align}
\alpha'f=\frac{4(m+2)(n+2)}{n-m}\left[(n+2)S'_{2n+1}(0)-(m+2)S'_{2m+1}(0)\right]\,.\label{alphaSR1b}
\end{align}
We have different expressions of sum rules for $f'$ and $\alpha'$ depending of the choice of $(m,n)$. We show explicitly how they are satisfied by string amplitudes in app.~\ref{sec:FESRconsitency}.\footnote{As a result, we can derive consistency conditions. For example, the relation $0=\eqref{alphaSR1a}|_{(m,n)=(0,2)}-\eqref{alphaSR1a}|_{(m,n)=(0,1)}$ gives rise to the following constraint:
\begin{align}
    0
    =
    \frac{1}{f'}
    \left\langle
		y\left[
			(48 y^5-40 y^4-36 y^3+27 y^2+4 y-2)
			-2 \J^2 \left(8 y^4-9 y^2+2\right)
		\right]
    \right\rangle_\uv
    \,. \label{consistency1}
\end{align}
The equation $0=\eqref{alphaSR1b}|_{(m,n)=(0,2)}-\eqref{alphaSR1b}|_{(m,n)=(0,1)}$ also provides the same result. The results shown in app.~\ref{sec:FESRconsitency} implies that this consistency condition also holds in string amplitudes. It would be interesting to study how this can be used to constrain the particle spectrum. We leave this aspect for our future work.}

Similarly, we can also derive sum rules for $\alpha''$ and $f''$ from FESRs for $S''_{2n+1}(0)$ as 
\begin{align}
S''_{2n+1}(0)=\frac{f''}{2(n+2)}-\frac{\alpha'}{n+2}S'_{2n+1}(0)-\frac{\alpha''f}{4(n+2)^2}\,.
\end{align}
Then, we can derive a sum rule for $\alpha''$ as
\begin{align}
	\frac{\alpha''f}{2}
	= -\alpha'f'
	+ 	\frac{\epsilon^2}{\Ms^4}
	\left\langle 
		y\left[a_0(y)+a_1(y)\J^2+a_2(y)\J^2(\J^2-2)\right]
	\right\rangle_\uv
	\,,\label{ddotalphabd1}
\end{align}
where 
\begin{subequations}
\label{defa}
\begin{align}
&a_0(y)=3360 y^5-4800 y^4-20 y^3+1944 y^2-402 y-56\,,\label{defa0}\\
&a_1(y)=-1920 y^4+1600 y^3+1296 y^2-972 y-112+56y^{-1}\,,\\
&a_2(y)= 160y^3-162y+28y^{-1}\,.
\end{align}
\end{subequations}
We can also derive different sum rules for $\alpha''$, although we do not show them here.

\subsection{Useful bounds on the amplitude in the low-energy regime}\label{sec:usefulbound}

Using FESRs \eqref{FESR1}, we can constrain the behavior of $\im\,\scat(s,t)$ in the low-energy regime $\Ms^2<s<\Mstar^2$, for which we cannot use the Regge behavior: 
the forward limit of \eqref{FESR1} gives 
\begin{align}
	S_{2n+1}(0)
	&=
	\left\langle
		y^{2n+2}
	\right\rangle_\uv
	=
	\frac{f}{2n+4}
	\qquad (n=0,1,2,\cdots)
	\,.\label{FESRfwd1}
\end{align}  
Since the condition $f\sim\Ms^4\alpha'\Mpl^{-2}\epsilon^{-2}$ is required by the absence of $t^{-1}$-singularity in \eqref{eq:disp2}, the condition \eqref{FESRfwd1} shows that the integral of $\rho_J(s)$ from $s=\Ms^2$ to $\Ms^2/\epsilon$ must be suppressed by $\Mpl^{-2}$. 
Eqs.~\eqref{FESRfwd1} can be also used to obtain the two-sided bounds on $\langle y^{1-k}\rangle_\uv$ with $k\geq0$: the strongest bound comes from \eqref{FESRfwd1} with $n=0$,
\begin{align}
	\frac{f}{4}
	\leq
	\left\langle
		y^{1-k}
	\right\rangle_\uv
	\leq
	\frac{f}{4\,\epsilon^{1+k}}
	\quad (k\geq0)
	\,.\label{originalpartial1}
\end{align}  
The $k=3$ case provides the two-sided bounds on the first term on the RHS of \eqref{eq:disp2} as 
\begin{align}
	\frac{\epsilon^2f}{\pi\Ms^4}
	\leq
	\frac{4}{\pi}\int^{\Ms^2/\epsilon}_{\Ms^2}\mathrm{d}s\,\frac{\im\,\scat(s,0)}{s^3}
	\leq
	\frac{f}{\pi\Ms^4\epsilon^2}
	\,.\label{boundfirst}
\end{align}

\paragraph{Examples.}
In the analysis below, we use eqs.~\eqref{alphaSR2a}, \eqref{ddotalphabd1}, and \eqref{FESRfwd1}. These are the FESRs for $f'$, $\alpha''$, and $f$, respectively. It is then useful to investigate the validity of these expressions in string amplitudes: see app.~\ref{sec:FESRconsitency} for details. It turns out that these FESRs are satisfied in a good approximation for sufficiently tiny $\epsilon\leq \mathcal{O}(0.01)\text{-}\mathcal{O}(0.1)$.

\section{Bounds from FESRs}\label{sec:FESRbound}
We derive bounds on the Regge parameters $f'$ and $\alpha''$ in sec.~\ref{sec:boundf'} and \ref{sec:boundalpha''}, respectively, by using the FESRs and the $\nullc_k$-constraints \eqref{nullapp}. We then derive bounds on $c_2$ in sec.~\ref{sec:boundc2}, proving \eqref{eq:scaling1}. We ignore the loop corrections from light particles in this section. A straightforward extension to the case where loops of light particles are included is discussed later in sec~\ref{sec:loop}.
\subsection{Constraints on $f'$}\label{sec:boundf'}
By using eqs.~\eqref{FESRfwd1}, \eqref{originalpartial1}, and an inequality $\langle y^{3}\rangle_\uv\leq\langle y^2\rangle_\uv$, we have a lower bound on $-f'$ from \eqref{alphaSR2a} as  
\begin{align}
	-f'
	>
	-\frac{7\epsilon f}{4\Ms^2}
	-
	\frac{\epsilon}{\Ms^2}
	\left\langle\,
			y(18y^2-8)\J^2
	\right\rangle_\uv	
	\,. \label{imrdotbd1}
\end{align}
To obtain a lower bound on the second term on the RHS, we use the $\nullc_k$-constraint \eqref{nullapp} 
to get
\begin{align}
	-f'
	>
	-\frac{7\epsilon f}{4\Ms^2}
	+
	\frac{\epsilon}{\Ms^2}
	\left\langle\,
			y^{n+1}\mathcal{I}_{n,k}(y,\J^2;\beta)
	\right\rangle_\uv	
	\,, \label{imdotint1}
\end{align}
where 
\begin{align}
	\mathcal{I}_{n,k}(y,\J^2;\beta)
	\coloneqq
	y^{-n}
	\left[
		(8-18y^2)\J^2
		+\beta y^{-5-k} \nullc_k(\J^2)
	\right]
	\,.\label{Idef}
\end{align}
We introduced an arbitrary positive constant $\beta$ and a non-negative integer $n$. Because $\beta \nullc_k(\J^2)\simeq 2\beta(\J^2)^k>0$ at large $J$, 
$\mathcal{I}_{n,k}(y,\J^2;\beta)$ is bounded from below by some constant $A_{n,k}(\beta)$ in the regions $\bigl\{(y,J):\,\epsilon\leq y\leq1\,,\,\, J\in\{0,2,4,\cdots\}\bigr\}$. In terms of $A_{n,k}(\beta)$, we can rewrite \eqref{imdotint1} as 
\begin{align}
	-f'
	>
	\frac{\epsilon}{\Ms^2}
		\left[
			-\frac{7f}{4}
			+
			A_{n,k}(\beta)
				\left\langle\,
					y^{1+n}
			\right\rangle_\uv
		\right]	
	\,. \label{imrdotint3}
\end{align}
Here, we used the unitarity condition $\rho_J\geq0$. The second term on the RHS can be evaluated by using \eqref{FESRfwd1} or \eqref{originalpartial1}. We can choose $\beta$ and $n$ to maximize $A_{n,k}(\beta)\langle y^{1+n}\rangle$. We set $n=1$ below for a while and explain why this choice will give the best bound. We however emphasize for definiteness that the bounds derived below are valid irrespective of whether the $n=1$ case provides the best bound or not.

Now we determine $A_{1,k}(\beta)$ and choose a positive free parameter $\beta$ appropriately to maximize $A_{1,k}(\beta)$. $\beta\nullc_k(\J^2)$ is nonnegative for $J\geq J_{k}+2\cdots$, while it can be negative for $J=0,2,4,\cdots J_k$. Here, the value of $J_k$ depends on $k$: {\it e,g.}, $J_3=J_4=4$, $J_5=J_6=6$, etc. Accordingly, the behavior of $\mathcal{I}_{1,k}(y,\J^2;\beta)$ in the small-$J$ region $0\leq J \leq J_k$ differs from the one in the large-$J$ region $J\geq J_k+2$. We discuss these two cases separately. Below, we consider $k=3,4,5,\cdots,24$, while we also consider $k=36$ case only in fig.~\ref{fig:plot_f'_tinyepsilon}.

\paragraph{Small $J$ analysis.}
We start with evaluating the minimum value of $\mathcal{I}_{1,k}(y,\J^2;\beta)$ within the region $y\in[\epsilon,1]$ and $J\in\{0,2,4,\cdots J_k\}$. For such small $J$, $\nullc_k(\J^2)$ can be negative. We define an even integer $J_{*,k}$ at which $\nullc_k(\J^2)$ becomes the smallest. Typically, we have $J_{*,k}=J_k$: we can check that this holds true for $k=3,4,\cdots, 24,36$ cases. $\mathcal{I}_{1,k}(y,\J^2;\beta)$ is dominated by the second term on the RHS of \eqref{Idef} when $\epsilon$ is sufficiently tiny and $\beta$ is not so small to satisfy $\beta\gtrsim\epsilon^{5+k}$. Then, for such $\epsilon$ and $\beta$, $\mathcal{I}_{1,k}(y,\J^2;\beta)$ will be minimized at $(y,J)=(\epsilon,J_{*,k})$:
\begin{align}
	A_{1,k}^{\text{small-$J$}}(\beta)
	\coloneqq
	\mathop{\min_{J\in\{0,2,\cdots,J_k\}}}_{y\in[\epsilon,1]}\mathcal{I}_{1,k}(y,\J^2;\beta)
	=
	\mathcal{I}_{1,k}(\epsilon,\J^2_{*,k};\beta)\,.\label{IlowJmin}
\end{align}
Assuming that \eqref{IlowJmin} is true, we choose  $\beta$ to optimize the bound. We will then confirm that eq.~\eqref{IlowJmin} is indeed valid for such $\beta$. $A_{1,k}^{\text{small-$J$}}(\beta)$ is a  monotonically decreasing function of $\beta$. 

\paragraph{Large $J$ analysis.}Next, we consider the $J\geq J_k+2$ case. We first use an inequality $8-18y^2\geq -10y$, which is valid within the region $y\in[\epsilon,1]$, to get 
\begin{align}
	\mathcal{I}_{1,k}(y,\J^2;\beta)
	\geq 
	-10\J^2+\beta y^{-6-k}\nullc_k(\J^2)
	\,. 
\end{align}
The RHS is minimized at $y=1$ because $\nullc_k(\J^2)>0$ for $J\geq J_k+2$. Also, we treat $J$ as continuous variables for convenience. Then, we have 
\begin{align}
	&\mathop{\min_{\text{even} J\geq J_k+2}}_{y\in[\epsilon,1]}\mathcal{I}_{1,k}(y,\J^2;\beta)
	\geq 
	A_{1,k}^{\text{large-$J$}}(\beta)
	\coloneqq
	\min_{J\geq J_k+2}
		\left[
			-10\J^2+\beta \nullc_k(\J^2)
		\right]
	\,.\label{IhighJmin}
\end{align}
The function $-10\J^2+\beta \nullc_k(\J^2)$ is a polynomial of $\J^2$ of degree $k$ and its minimum can be evaluated analytically in principle. $-10\J^2+\beta \nullc_k(\J^2)$ is a monotonically increasing function of $\beta$, as well as $A_{1,k}^{\text{large-$J$}}(\beta)$. 

\begin{figure}[tbp]
 \centering
  \includegraphics[width=.7\textwidth, trim=150 180 200 160,clip]{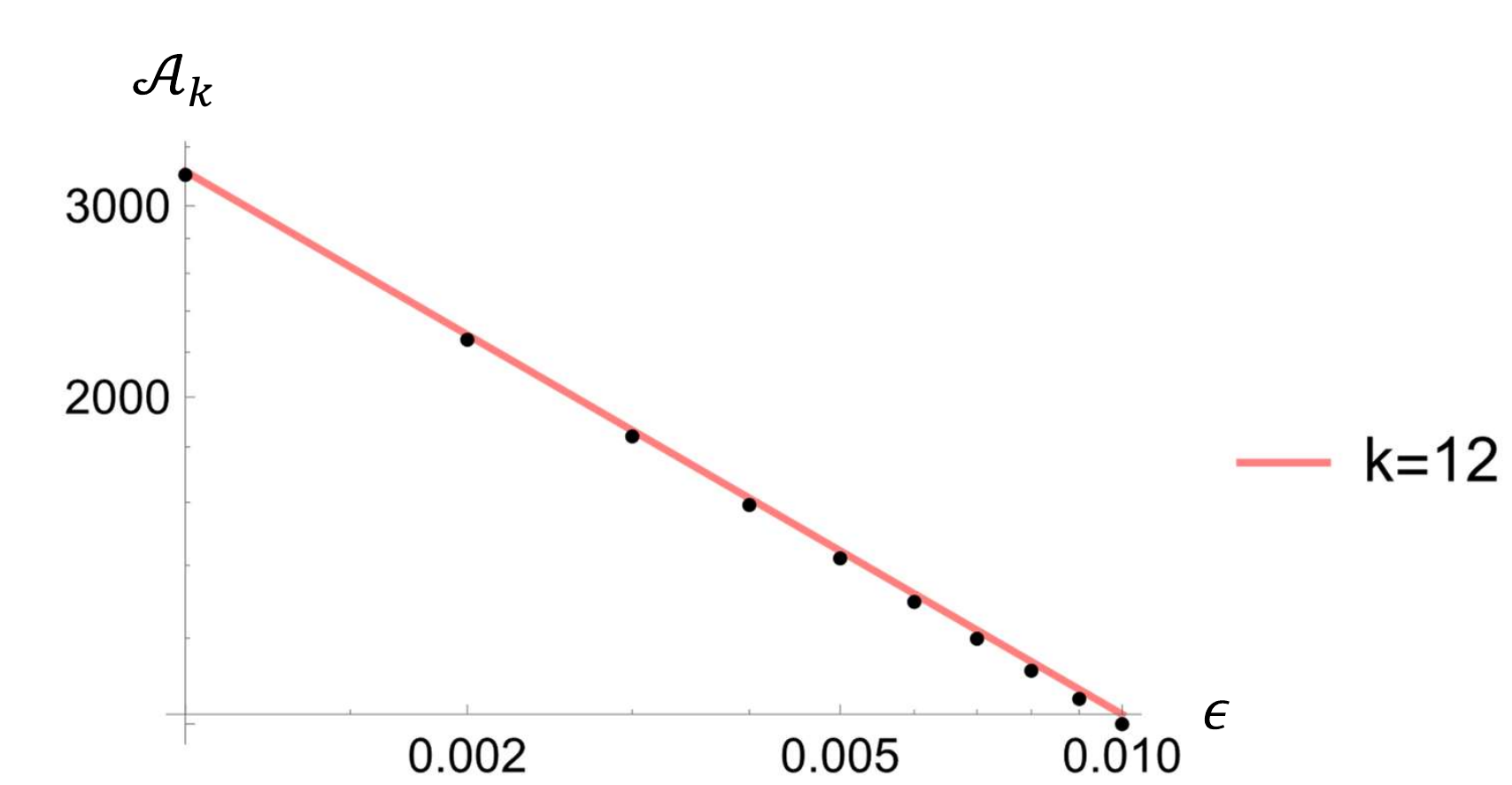}
 \caption{A plot of $\mathcal{A}_k$ with $k=12$. Black dots are numerical results obtained from $\beta_k^\text{num}$ with $\epsilon=0.001,0.002,\cdots,0.01$. The solid line expresses analytic estimates \eqref{Akapp} with $k=12$ which are valid at the leading order in small-$\epsilon$ expansions. This plot shows the analytic result correctly reproduces the numerical results in a good approximation.}
 \label{f'_num_vs_ana} 
\end{figure}

\paragraph{Derivation of $A_{1,k}(\beta)$.}
We can then derive $A_{1,k}(\beta)$ from eqs.~\eqref{IlowJmin} and \eqref{IhighJmin} via $
A_{1,k}(\beta)=\min\bigl[A_{1,k}^{\text{small-$J$}}(\beta),\,A_{1,k}^{\text{large-$J$}}(\beta)\bigr]$. 
Because $A_{1,k}^{\text{small-$J$}}(\beta)$ and $A_{1,k}^{\text{large-$J$}}(\beta)$ are monotonically increasing and decreasing functions of $\beta$, respectively, there exists a point $\beta=\beta_k^\text{exact}$ for which we have 
\begin{align}
	A_{1,k}^{\text{small-$J$}}(\beta_k^\text{exact})
	= A_{1,k}^{\text{large-$J$}}(\beta_k^\text{exact})
	\,.\label{beta_exact}
\end{align}
Consequently, $A_{1,k}(\beta)$ is maximized at $\beta=\beta_k^\text{exact}$. We solve eq.~\eqref{beta_exact} numerically for given $\epsilon$. Writing the numerical solution for given $\epsilon$ as $\beta_k^\text{num}$, we compute $A_{1,k}(\beta_k^\text{num})$. In terms of the obtained $A_{1,k}(\beta_k^\text{num})$, the bound \eqref{imrdotint3} reads
\begin{align}
	-f'/f
	>
	-\frac{\mathcal{A}_{k}}{\Ms^2}
	\,,\qquad
	\mathcal{A}_{k}
	\coloneqq \frac{1}{4}
		\left[
			-\epsilon A_{1,k}(\beta_k^\text{num})+7\epsilon
		\right]
	\,.
 \label{f'bound1}
\end{align}
Here, we used \eqref{FESRfwd1}. As an illustrative example, we consider the $k=12$ case with $\epsilon=0.001,0.002,\cdots, 0.01$. We can check numerically that eq.~\eqref{IlowJmin} is indeed valid for our choice of $(\epsilon,\beta_k^\text{num})$.
The value of $\mathcal{A}_k[\beta_k^\text{num}]$ as a function of $\epsilon$ is plotted in fig.~\ref{f'_num_vs_ana}.  

We do not perform numerical computations for $k>12$ cases in the present analysis because the numerical computations become heavier for higher-$k$. For such higher-$k$ cases, we use the analytic expressions which will be derived soon in sec.~\ref{sec:analyticf}.

\paragraph{Comments on $n\neq1$ cases.} Let us briefly consider the $n\neq1$ cases.  For $n=0$ case, we get an additional $\epsilon^{-1}$ factor when using \eqref{originalpartial1} to estimate $\langle y \rangle_\uv$ in \eqref{imrdotint3}. We do not have this enhancement when estimating $\langle y^{1+n}\rangle_\uv$ with $n\geq1$. Due to this, the bound in the $n=1$ case will be stronger than those obtained in the $n=0$ case when $\epsilon\ll1$. For $n>1$ cases, we have $A_{n,k}(\beta)=\min\bigl[A_{n,k}^{\text{small-$J$}}(\beta),\,A_{n,k}^{\text{large-$J$}}(\beta)\bigr]$ with 
$A_{n,k}^{\text{small-$J$}}(\beta)=\mathcal{I}_{n,k}(\epsilon,\J^2_{*,k};\beta)$ and $A_{n,k}^{\text{large-$J$}}(\beta)=\min_{J\geq J_k+2}\bigl[-a_n\J^2+\beta \nullc_k(\J^2)\bigr]$, analogously to eqs.~\eqref{IlowJmin} and \eqref{IhighJmin}.  Here, $a_n\coloneqq \min_{y\in[\epsilon,1]}[(8-18y^2)y^{-n}]$. We have $A_{n,k}^\text{large-$J$}\leq A_{1,k}^\text{large-$J$}(<0)$ because $a_n\leq -10$ for $n>1$. Also, we have $A_{n,k}^\text{small-$J$}\ll A_{1,k}^\text{small-$J$}(<0)$ for $\epsilon\ll1$ because $\mathcal{I}_{n,k}/\mathcal{I}_{1,k}\sim\epsilon^{1-n}$. Hence, we will have $A_{n,k}(\beta)\ll A_{1,k}(\beta)(<0)$ and consequently the bounds in the $n>1$ cases will be weaker than the one obtained in the $n=1$ case when $\epsilon\ll1$.

\subsubsection{Analytic expression}\label{sec:analyticf}
It is useful to derive an analytic expression of $\mathcal{A}_k$ at the lowest order in small $\epsilon$. For this purpose, we derive an analytic expression of $A_{1,k}^\text{large-$J$}(\beta)$ which is valid at the leading order in small $\epsilon$-expansions. When $\beta$ is sufficiently small, the minimum of $-10\J^2+\beta \nullc_k(\J^2)$ is realized at $J\gg1$. In this case, we can approximate the function around its minimum as 
\begin{align}
	-10\J^2+\beta \nullc_k(\J^2)
	\simeq
	-10\J^2+2\beta(\J^2)^k
	\,,
\end{align}
where the RHS is minimized at $\J^2=(\beta k/5)^{\frac{1}{1-k}}\gg1$ for sufficiently small $\beta$. We then obtain a simple but approximate expression of $A_{1,k}^\text{large-$J$}(\beta)$ as
\begin{align}
	A_{1,k}^\text{large-$J$}(\beta)
	\simeq
	\frac{-10(k-1)}{k}
		\left(
			\frac{5}{\beta k}
		\right)^{\frac{1}{k-1}}
	\,.\label{IhighJminapp}
\end{align}
We can also approximate $A_{1,k}^\text{small-$J$}$ as 
\begin{align}
	A_{1,k}^\text{small-$J$}
	\simeq 
	\beta\epsilon^{-6-k} \nullc_k(\J^2_{*,k})
	\,.\label{IlowJminapp}
\end{align}
We approximately solve $A_{1,k}^\text{small-$J$}(\beta)=A_{1,k}^\text{large-$J$}(\beta)$ by using \eqref{IhighJminapp} and \eqref{IlowJminapp} to get an approximate expression of $\beta_k^\text{exact}$: the result is 
\begin{align}
	\beta_k^\text{exact}
	\simeq
	\beta_k^\text{approx}
	\coloneqq
	\frac{5}{k}
		\left(
			\frac{-\nullc_k(\J^2_{*,k})}{2k-2}
		\right)^{-1+\frac{1}{k}}
	\epsilon^{5+k-\frac{6}{k}}
	\,.\label{beta_app}
\end{align}
This shows $\beta_k^\text{approx}\gg\epsilon^{5+k}$ when $\epsilon$ is sufficiently tiny, providing an analytic understanding why eq.~\eqref{IlowJmin} is indeed valid when choosing $\beta\sim\beta_k^\text{exact}$. Substituting \eqref{beta_app} into $A_{1,k}^\text{small-$J$}(\beta)$ or $A_{1,k}^\text{large-$J$}(\beta)$, we obtain an approximate expression of $A_{1,k}(\beta)$. Finally, an approximate expression of $\mathcal{A}_k$ is obtained as 
\begin{align}
	\mathcal{A}_k
	\simeq
	\frac{5(k-1)}{2k}
		\left(
			\frac{-\nullc_k(\J^2_{*,k})}{2k-2}
		\right)^{\frac{1}{k}}
	\epsilon^{-\frac{6}{k}}
	\,.\label{Akapp}
\end{align}
\begin{figure}[tbp]
 \centering
  \includegraphics[width=.7\textwidth, trim=120 160 160 130,clip]{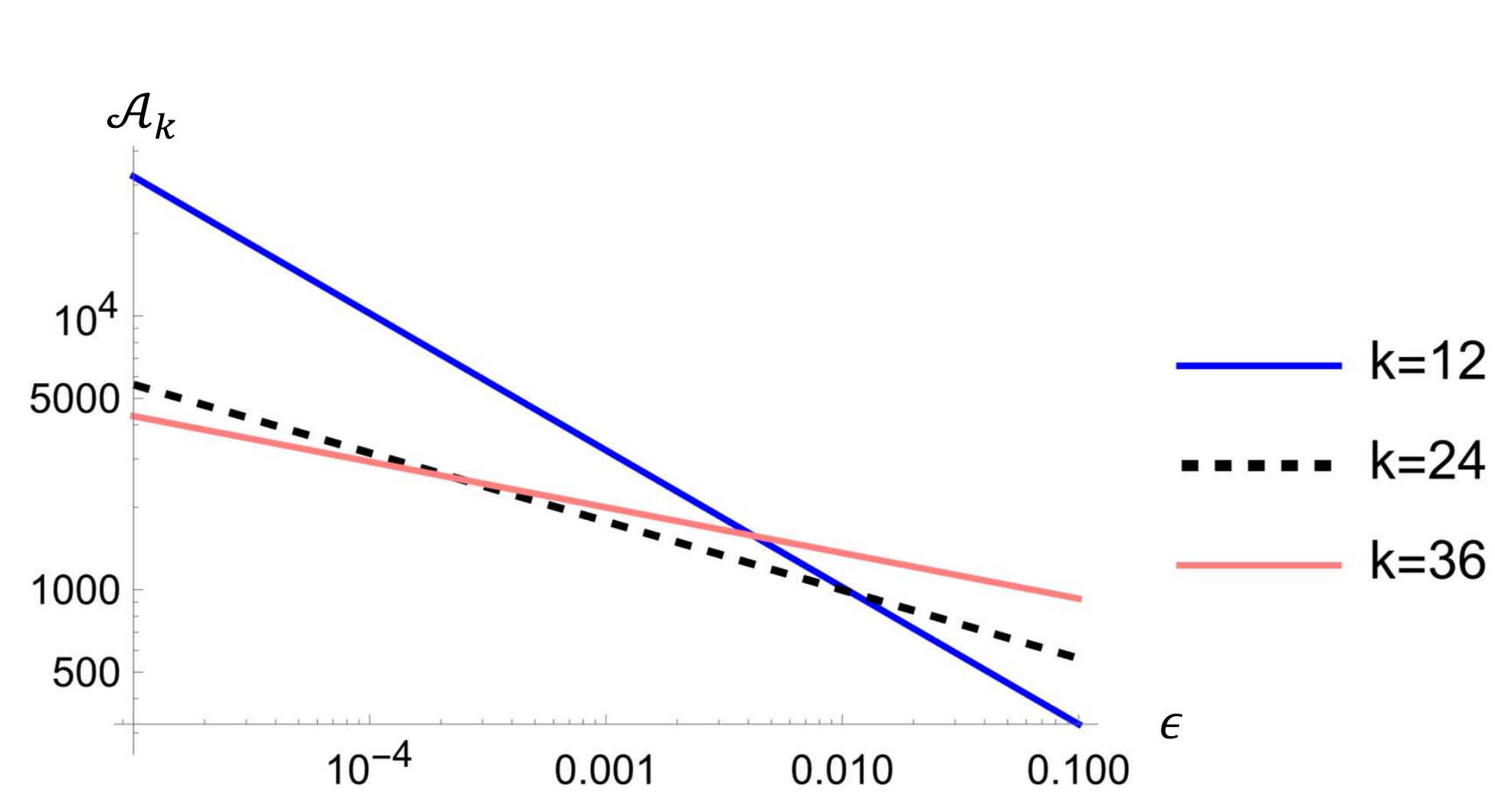}
 \caption{A plot of $\mathcal{A}_k$ with $k=12,24,36$ based on \eqref{Akapp}. A region $10^{-5}\leq\epsilon\leq 10^{-1}$ is considered. This figure shows that large-$k$ results provide better bounds for smaller $\epsilon$.}
 \label{fig:plot_f'_tinyepsilon} 
\end{figure}
The analytic estimates \eqref{Akapp} are compared with the numerical results in fig.~\ref{f'_num_vs_ana} for $k=12$ case. The figure indicates that the analytic estimates \eqref{Akapp} correctly reproduce the numerical results in a good approximation. Since the numerical evaluations of $\mathcal{A}_k$ become heavier for higher-$k$ cases, eqs.~\eqref{Akapp} are useful to derive the bounds for such higher-$k$ efficiently. In the present study, we use \eqref{Akapp} for evaluating $\mathcal{A}_k$ with $k>12$. The results \eqref{Akapp} suggest that the large $k$ cases provides the better bound when considering very tiny $\epsilon$. For example, fig.~\ref{fig:plot_f'_tinyepsilon} shows that for $\epsilon\lesssim 10^{-2} (10^{-4})$, the bound with $k=24$ $(k=36)$ is much better than the one with $k=12$. 

\begin{figure}[tbp]
 \centering
  \includegraphics[width=.7\textwidth, trim=150 180 210 160,clip]{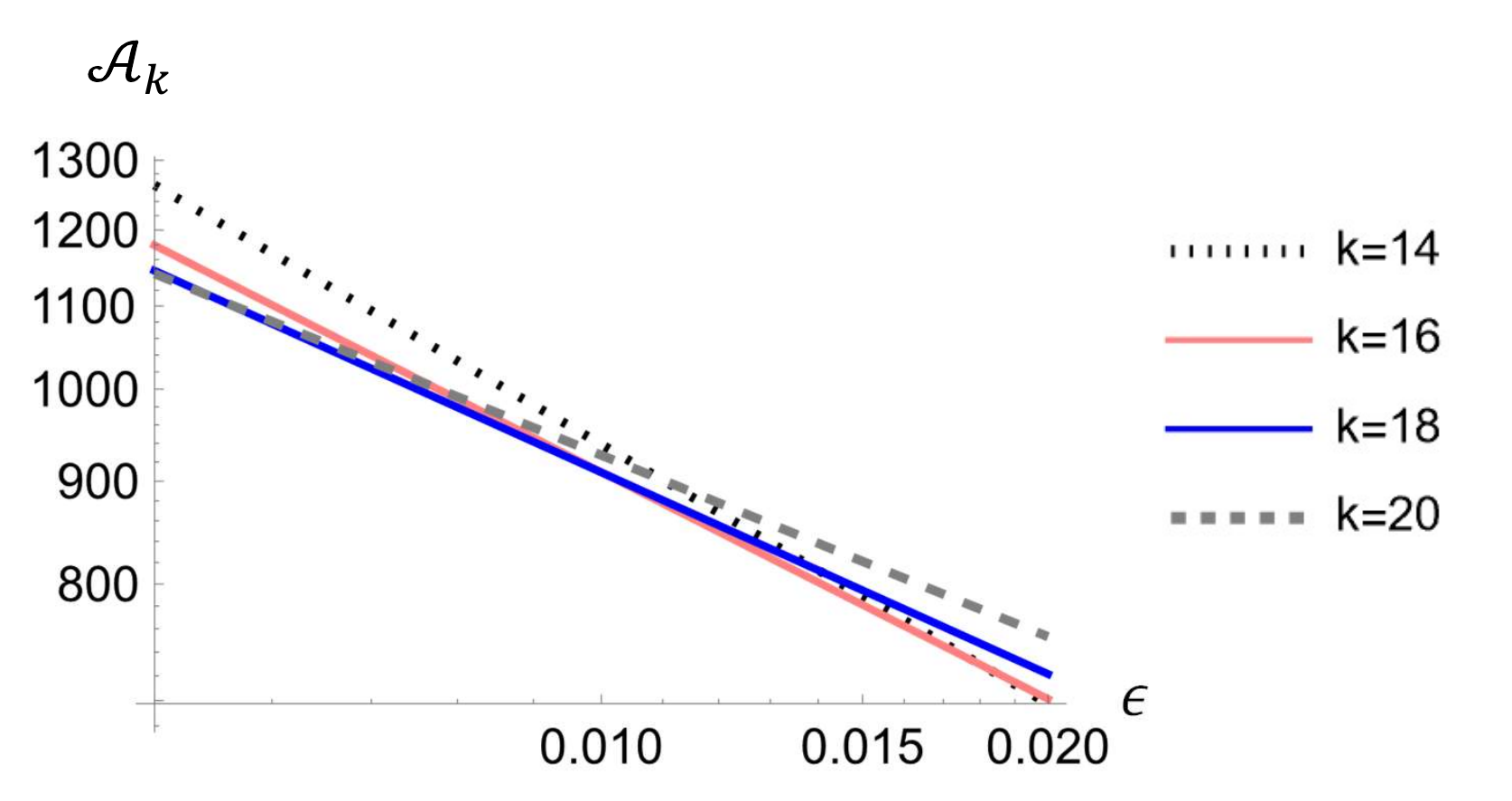}
 \caption{A plot of $\mathcal{A}_k$ with $k=14,16,18,20$, based on \eqref{Akapp}. The $k=14,16,18,20$ cases are expressed by the black dotted line, the pink solid line, the blue solid line, and the gray dashed line, respectively. 
 The region $0.005\leq\epsilon\leq0.02$, including the benchmark point $\epsilon=0.01$, is considered here. In this region, the bounds from $k=14,16,18,20$ give the tightest lower bound on $-f'/f$ via \eqref{f'bound1}.} 
 \label{fig:plot_f'} 
\end{figure}

By using \eqref{Akapp}, we find that the bounds from $k=14,16,18,20$ give the tightest lower bound on $-f'/f$ via \eqref{f'bound1} when considering the region $0.005\leq\epsilon\leq0.02$ in which the benchmark point $\epsilon=0.01$ is included: see fig.~\ref{fig:plot_f'}.

\subsection{Constraints on $\alpha''$}\label{sec:boundalpha''}
\begin{figure}[tbp]
 \centering
  \includegraphics[width=.67\textwidth, trim=130 200 240 170,clip]{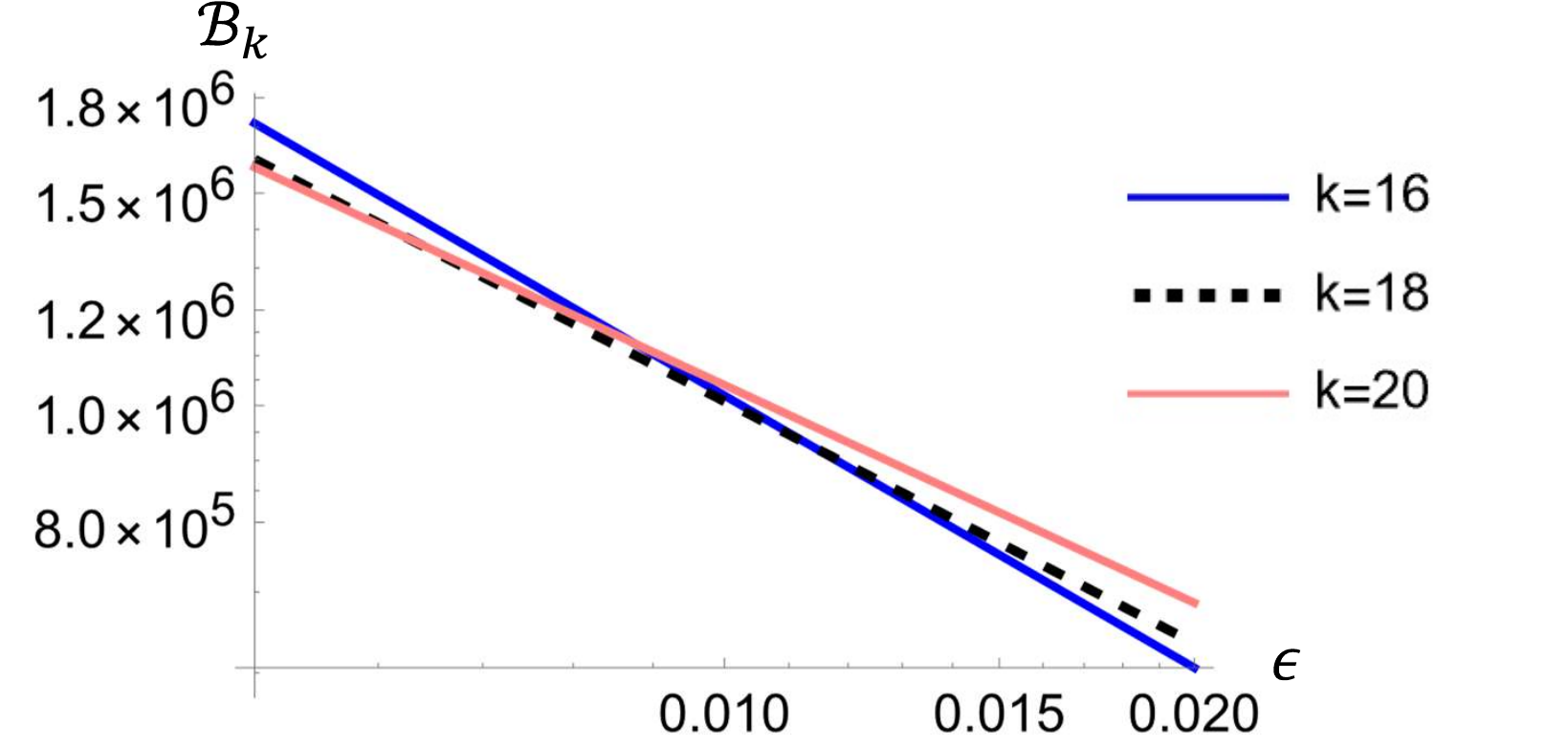}
 \caption{A plot of $\mathcal
{B}_k$ with  $k=16,18,20$, based on \eqref{Bkapp}. The $k=16,18,20$ cases are expressed by the blue solid line, the black dashed line, and the pink solid line, respectively. The region $0.005\leq\epsilon\leq0.02$, including the benchmark point $\epsilon=0.01$, is considered here. In this region, the bounds from $k=16,18,20$ give the tightest lower bound on $\alpha''/\alpha'$ via \eqref{alpha''bound1}. }
 \label{fig:plot_alpha''} 
\end{figure}

Next, we derive a lower bound on $\alpha''$ from the sum rule ~\eqref{ddotalphabd1} for $\alpha''$. By using an inequality $a_0(y)>3360 y^5-2876y^3 - (402+56/\epsilon) y$, which is valid for $y\in(\epsilon,1)$, we obtain a lower bound on $\alpha''$ from \eqref{ddotalphabd1} as
\begin{align}
	\frac{\alpha''f}{2}
	>
	& - \alpha'f'
	- \frac{f}{\Ms^4}
			\left(
				14\epsilon+\frac{959\epsilon^2}{6}
			\right)
	+ \frac{\epsilon^2}{\Ms^4}
		\left\langle
			y^{1+n}\mathcal{H}_{n,k}(y,\J^2;\gamma)
		\right\rangle_\uv
	\,,\label{ddotalphabd2}
\end{align} 
where $\mathcal{H}_{n,k}(y,\J^2;\gamma)$ is defined by 
\begin{align}
	\mathcal{H}_{n,k}(y,\J^2;\gamma)
	\coloneqq 
	y^{-n}
		\left[
			a_1(y)\J^2+a_2(y)\J^2(\J^2-2)+\gamma y^{-5-k} \nullc_k(\J^2)
		\right]
	\,.
    \label{Hdef}
\end{align}
To obtain \eqref{ddotalphabd2}, we also used the $\nullc_k$-constraint \eqref{nullapp}. We introduced a positive parameter $\gamma$ and a nonnegative integer $n$. Because $\gamma\nullc_k(\J^2)\simeq2\gamma(\J^2)^k>0$ at large $J$, $\mathcal{H}_{n,k}(y,\J^2;\gamma)$ is bounded from below  by some constant $B_{n,k}(\gamma)$ in the regions $\bigl\{(y,J):\,\epsilon\leq y\leq1\,,\,\, J\in\{0,2,4,\cdots\}\bigr\}$. In terms of $B_{n,k}(\gamma)$, we can rewrite \eqref{ddotalphabd2} as 
\begin{align}
	\frac{\alpha''f}{2}
	>
	& - \alpha'f'
	- \frac{f}{\Ms^4}
			\left(
				14\epsilon+\frac{959\epsilon^2}{6}
			\right)
	+ \frac{\epsilon^2B_{n,k}(\gamma)}{\Ms^4}
		\left\langle
			y^{1+n}
		\right\rangle_\uv
	\,.\label{ddotalphabd3}
\end{align} 
Here, we used the unitarity condition $\rho_J\geq0$. We can choose $n$ and $\gamma$ to optimize the bound. The methodology to derive a lower bound on $\alpha''$ from \eqref{ddotalphabd3} is completely the same as the one discussed in sec.~\ref{sec:boundf'}. Hence, we show the final result only in the main text, while the detailed analysis is performed in app.~\ref{sec:boundalpha''detail}: the result is
\begin{align}
\alpha''/\alpha'
	>
	& - 2f'/f
	- \frac{2\mathcal{B}_k}{\Ms^4\alpha'}
	\,.\label{alpha''bound1}
\end{align}
As is done in sec.~\ref{sec:boundf'}, we can also derive an approximate analytic expression of $\mathcal{B}_k$ at the leading order in small $\epsilon$-expansions as
\begin{align}
	\mathcal{B}_k
	\simeq
        \frac{81-8\sqrt{70}}{2}
	\frac{k-2}{k}
		\left(
			\frac{-\nullc_k(\J^2_{*,k})}{k-2}
		\right)^{\frac{2}{k}}
	\epsilon^{-\frac{12}{k}}
	\,.\label{Bkapp}
\end{align}
By using \eqref{Bkapp}, we find that the bounds from $k=16,18,20$ give the tightest lower bound on $\alpha''/\alpha'$ via \eqref{alpha''bound1} when considering the region $0.005\leq\epsilon\leq0.02$ in which the benchmark point $\epsilon=0.01$ is included: see fig.~\ref{fig:plot_alpha''}.

\subsection{Constraints on $c_2$} \label{sec:boundc2}
Now we derive a lower bound on $c_2(0)$ by substituting eqs.~\eqref{f'bound1} and \eqref{alpha''bound1} into \eqref{eq:disp3} as\footnote{To improve a lower bound on $c_2(0)$, we may also need to evaluate the first term on the RHS of \eqref{eq:disp2} which is ensured to be positive. We have a two-sided bound \eqref{boundfirst}. However, this two-sided bound is too weak to improve the lower bound \eqref{F0bound} by taking into account the relation $f\sim\Ms^4\alpha'\Mpl^{-2}\epsilon^{-2}$. We thus simply use \eqref{eq:disp3}.} 
\begin{align}
    c_2(0)
    >
	F_0
	&>
	\frac{-1}{\Mpl^2\Ms^2}
    \left\{
		\min_{k=3,4,\cdots}
        \left[
			4\mathcal{A}_k
		\right]
    +
    \min_{k=3,4,\cdots}
        \left[
			\frac{2\mathcal{B}_k}{\Ms^2\alpha'}
		\right]
    -2\epsilon
    \right\}
   	\label{F0bound}
\end{align}
with \eqref{Akapp} and \eqref{Bkapp}. This is one of the main result of this paper. For smaller $\epsilon$, the best bound is given by the larger $k$ as illustrated in fig.~\ref{fig:plot_f'_tinyepsilon}. 
Since the $\epsilon$-dependence becomes weaker for large $k$, our bound \eqref{F0bound} is robust. For example, even for a wide range of $\epsilon$ with $10^{-5}\lesssim\epsilon\lesssim 10^{-2}$, we have 
$10^3\lesssim \mathcal{A}_k\lesssim 5\times 10^3$ and $10^6\lesssim \mathcal{B}_k\lesssim 2\times 10^7$.
As a benchmark point, we choose $\epsilon=0.01$. In this case, the best bound on $c_2(0)$ is obtained by taking $k=18$:
\begin{align}
    c_2(0)
    >
    F_0
    >
  \frac{-1}{\Mpl^2\Ms^2}
		\left(
			3.7\times 10^3
			+\frac{2.0\times 10^6}{\Ms^2\alpha'}
		\right)
	\,.\label{F0boundnum}
\end{align}
This proves \eqref{eq:scaling1}, that the unknown scale $M$ is bounded by $\Ms$, the mass of lightest higher-spin states. Physically, this bound states that the contributions from tower of higher-spin particles to $c_2(0)$ can be negative, while its negativity must be suppressed not only by $\Ms^{-1}$ but also by $\Mpl^{-1}$, at least when loops of light particles are ignored. Thanks to this suppression, our bound \eqref{F0boundnum} can be regarded as the approximate positivity bound which can be used to constrain the parameter space of gravitational EFTs in four spacetime dimensions. This suppression of negativity has been proposed qualitatively in \cite{Hamada:2018dde}, and our result verifies it quantitatively at least for the four-point amplitude of identical massless scalar.  It would be straightforward to extend our formalism to generic processes such as the light-by-light scattering. We leave this aspect for future work. 

\paragraph{Discussions.}
Our quantitative bound \eqref{F0boundnum} is much weaker than the bound expected from the EFT power counting $c_2(0)>-\mathcal{O}(1)/(\Mpl^{2}\Ms^{2})$, however. We can obtain stronger bounds on $c_2(0)$ in higher dimensions $D>4$: 
we only show the results here, and details of a straightforward extension of our formalism to higher dimensions can be found in app.~\ref{sec:general_d}. Let us parameterize bounds on 
$f'/f$, $\alpha''/\alpha'$, and $c_2(0)$ in general $D$-dimensions analogously to \eqref{f'bound1}, \eqref{alpha''bound1}, and \eqref{F0bound} as 
\begin{align}
       & -f'/f>-\frac{\mathcal {A}_k^{(D)}}{\Ms^2}
    \,,\qquad
    \alpha''/\alpha'
    >
    -2f'/f-\frac{2\mathcal {B}_k^{(D)}}{\Ms^4\alpha'}
    \,,\label{reggebound_d}\\
     &c_2(0)
	>
	\frac{-1}{\Mpl^2\Ms^2}
    \left\{
		\min_{k=3,4,\cdots}
        \left[
			4\mathcal{A}_k^{(D)}
		\right]
    +
    \min_{k=3,4,\cdots}
        \left[
			\frac{2\mathcal{B}_k^{(D)}}{\Ms^2\alpha'}
		\right]
    -2\epsilon
    \right\}
    \,.	\label{F0bound_d}
\end{align}
The values of $\mathcal{A}_k^{(D)}$ and $\mathcal{B}_k^{(D)}$ are estimated at the benchmark point $\epsilon=0.01$ as\footnote{Note that the minimum of $\mathcal{A}_k^{(D)}$ at $\epsilon=0.01$ are realized at $k=18$ and $k=20$ for $D=4,5,\cdots,10$ and $D=11,12$, respectively. The minimum of $\mathcal{B}_k^{(D)}$ at $\epsilon=0.01$ are realized at $k=18$ and $k=20$ for $D=4,5,\cdots,8$ and $D=9,10,11,12$, respectively. }  
\begin{subequations}
\label{akbkvalue_d}
\begin{align}
    &\min_{k=3,4,\cdots}\left[\mathcal{A}_k^{(D)}\right]_{\epsilon=0.01}
    \simeq
    10^2\times \{9.1,6.4,5.1,4.3,3.7,3.4,3.1,2.8,2.7\}
    \,\\
    &\min_{k=3,4,\cdots}\left[\mathcal{B}_k^{(D)}\right]_{\epsilon=0.01}
    \simeq
    10^5\times\{10,6.0,4.2,3.2,2.6,2.1,1.8,1.6,1.4\}
\end{align}
\end{subequations}
for $D=4,5,\cdots,12$. Our bounds on $c_2(0)$ in higher dimensions are still much weaker than the bound expected from the power counting. Hence, the weakness of our bound will not be a feature peculiar to the $D=4$ case. We expect the existence of the theoretical bounds which are much stronger than our bounds in general dimensions.

Indeed, such stronger bounds have been derived in higher dimensions $D>4$ in \cite{Caron-Huot:2021rmr} by using the twice-subtracted dispersion relation in the finite impact parameter space $b\sim \Ms$. The Reggeization of the graviton exchange is not assumed there. Consequently, the contribution from the graviton $1/t$-pole explicitly remains in their sum rules. Their bounds in higher dimensions $D>4$ are,\footnote{Ref.~\cite{Caron-Huot:2021rmr} derived the bound on $g_2$, where $g_2$ is related to our $c_2(0)$ as $c_2(0)=2g_2$.} 
\begin{align}
    &c_2(0)|_{\text{Ref.~\cite{Caron-Huot:2021rmr}}}
    >
    -\frac{1}{\Mpl^2\Ms^2}\times \{36,19,14,11,9.6,8.5,7.8,7.4\}
    \qquad
    (D=5,6,\cdots,12)
    \,,\label{simon_bound}
\end{align}
where we calculated the numerator to two significant digits based on the set of inequalities given in the table 3 of \cite{Caron-Huot:2021rmr}. The bounds \eqref{simon_bound} are close to those expected from the power counting and much stronger than our bounds in the $D>4$ case. 

In $D=4$, however, the bound obtained in \cite{Caron-Huot:2021rmr} is IR divergent:
\begin{align}
    c_2(0)|_{\text{Ref.~\cite{Caron-Huot:2021rmr}}}
    >
    -\frac{50}{\Mpl^2\Ms^2}\log(0.3\Ms b_\text{max})
    \qquad
    (D=4)
    \,,
\end{align}
where  $b_\text{max}$ denotes an IR cutoff introduced in the impact parameter space. This is because, the presence of graviton $1/t$-pole in their sum rule obstructs the derivation of the twice-subtracted dispersion relation in the finite impact parameter space in $D=4$ without introducing an IR cutoff. 
By contrast, our bound \eqref{F0boundnum} in $D=4$ is manifestly IR finite. This is because the problematic graviton $1/t$-pole does not appear in our sum rule \eqref{eq:disp2} of $c_2(0)$: it is cancelled with the contributions from the gravitational Regge amplitude as demonstrated in sec.~\ref{sec:review}. In $D=4$ flat background, the quantitative bound on $c_2(0)$ which is IR finite is only our bounds \eqref{F0boundnum} to our knowledge. We also note that our bounds on the Regge parameters $f'/f$ and $\alpha''/\alpha'$ are new in the literature in general dimensions $D\geq 4$. A price we pay in our approach is the additional assumption of the Reggeization of graviton exchange.

\subsection{Comparison with the type II superstring amplitude}
We can compare our bounds on $c_2(0)$, $f'/f$, and $\alpha''/\alpha'$ with concrete amplitudes. As an illustrative example, we consider the type II closed superstring amplitude. The relevant Regge parameters for this amplitude is given in eq.~\ref{VSreggeparameter}. Also  $c_2(0)=0$ as a consequence of supersymmetry. We summarize these results:
\begin{align}
    c_2(0)=0\,,\qquad
    f'/f=\frac{1}{\Ms^2}\left[2\gamma+2\ln(1/\epsilon)+\epsilon\right]\,,\qquad \alpha'=\Ms^2/2\,,\qquad \alpha''=0\,,
\end{align}
where $\gamma$ denotes the Euler-Mascheroni constant.
Our bounds on $c_2(0)$ and $\alpha''/\alpha'$ are trivially satisfied in this example. As for the $f'/f$, we need to specify the value of $\epsilon$. The analysis in app.~\ref{sec:FESRconsitency} implies that we can set $\epsilon=0.1$ with maintaining the validity of FESRs in a good approximation. With this choice, the value of $f'/f$ in this example is $f'/f|_{\epsilon=0.1}\simeq 5.86\Ms^{-2}$ while
our bounds on $f'/f$ in general dimensions read 
\begin{align}
    f'/f
    < 
    \min_{k=3,4,\cdots}\left[\frac{\mathcal{A}_k^{(D)}}{\Ms^2}\right]_{\epsilon=0.1}
    \simeq
    \frac{10^{2}}{\Ms^2}\times
    \{3.0,2.2,1.8,1.5,1.4,1.3,1.2,1.1,1.0\}
    \,,\label{fpbound}
\end{align}
for $D=4,5,\cdots,12$. Note that the best bounds for $\epsilon=0.1$ are obtained by taking $k=8$ and $k=10$ for $D=4$ and $D=5,6,\cdots,12$, respectively. We find that our bounds are easily satisfied in the type II amplitude. It would be interesting to study if one can sharpen the bounds such that they are saturated by some string amplitude, leaving this direction for our future work.

\section{Remarks}\label{sec:remark}
In this section, we summarize several remarks. In sec.~\ref{sec:FESRgravposi}, we stress that the gravitational positivity bounds \eqref{eq:disp2} or \eqref{eq:disp3} can be formulated as an FESR. In sec.~\ref{sec:subleading}, we briefly discuss how our FESRs are affected by the sub-leading terms which are suppressed in \eqref{eq:regge1} or \eqref{spin-2domi}. In sec.~\ref{sec:loop}, we discuss how our analysis can be extended to the case where loops of light particles are included. In sec.~\ref{sec:assumptions}, we summarize the assumptions we imposed in the present analysis.

\subsection{Gravitational positivity bounds as an FESR}\label{sec:FESRgravposi}
In sec.~\ref{sec:FESR}, we derived FESRs for the Regge parameters, starting from \eqref{start1}. In \eqref{start1}, we considered the complex integral of $\scat(s,t)(s+(t/2))^{2n+1}$ along the contour $\mathcal{C}_++\mathcal{C}_L$ with $n=0,1,2,\cdots$. If we do the same analysis with $n\leq -1$, then we will have additional terms on the RHS of \eqref{start1} and consequently on \eqref{FESRorigin}. In particular, for $n=-2$, we have
\begin{align}
	\frac{-2}{\Mpl^2t}+c_2(t)
	+\frac{4}{\pi}\left(\frac{\Ms^2}{\epsilon}+\frac{t}{2}\right)^{-2}\frac{f(t)}{\alpha(t)-2}
	=
	\left(\frac{\Ms^2}{\epsilon}+\frac{t}{2}\right)^{-2}\frac{4}{\pi}S_{-3}(t)\,.\label{gravposiFESR1}
\end{align}
Because the terms other than the first and the third terms on the LHS are regular in the vicinity of $t=0$, \eqref{gravposiFESR1} requires the cancellation of $t^{-1}$ term on the LHS. We then find that \eqref{gravposiFESR1} reduces to \eqref{eq:disp2} in the forward limit.  
The expression \eqref{gravposiFESR1} is valid even for $t>0$, while it is identical to the relation derived from the twice-subtracted dispersion relation for $t<0$. The essentially important assumption here is only the dominance of the Regge pole with the spin-$2$ Regge intercept \eqref{spin-2domi}.

An advantage of understanding the gravitational positivity bound \eqref{eq:disp2} as an FESR \eqref{gravposiFESR1} is that it becomes manifest that 
only the information of $\scat$ below the Reggeization scale $\Mstar^2$ is relevant for constraining $c_2(t)$. As long as there is a hierarchy between $\Ms$ and $\Mpl$ such that $\Ms\ll\Mpl$, {\it i.e.,} if gravity is UV completed within its weakly-coupled regime, it is not necessary to concern about the super-Planckian physics where strong dynamics of gravity such as creation of black holes and baby universes may be important.

\subsection{Influence of sub-leading corrections} \label{sec:subleading}
In general, we have sub-leading terms which are ignored in the approximation \eqref{eq:regge1} or \eqref{spin-2domi}. Let us write such terms as $\scat(\Ms^2/\epsilon,t)=\scat_\text{R}(\Ms^2/\epsilon,t)+\scat_\text{sub}(\Ms^2/\epsilon,t)$. $\scat_\text{sub}$ can be the contributions from daughter trajectories, for instance. The presence of such term may modify the FESRs for $f'$ and $\alpha''$ derived in sec.~\ref{sec:FESR}, even though we expect that such contributions can be ignored for sufficiently tiny $\epsilon$. In fact, we can disentangle such contributions once the high-energy behavior of such terms is fixed. We can then derive FESRs for $f'$ and $\alpha''$ by considering suitable linear combinations of $\{S_{2n+1}\}$. As the simplest toy example to demonstrate this, we consider the sub-leading term of the form $\im\,\scat_\text{sub}(s,t)\simeq g(t)\bigl(s+(t/2)\bigr)\bigl(\Ms^2/\epsilon+(t/2)\bigr)^{-1}$. 
In this case, eq.~\eqref{FESR1} is modified as
\begin{align}
	S_{2n+1}(t)
	=
	\frac{f(t)}{\alpha(t)+2n+2}
    +
    \frac{g(t)}{2n+3}
     \qquad 
	(n=0,1,2,\cdots)
	\,.\label{FESR1mod}
\end{align}
In this case, the LHS of FESR \eqref{alphaSR2a} is replaced by $f'+\frac{14}{15}g'$, where $g'\coloneqq\der_t g(t)|_{t=0}$. In this case, however, we consider $-204 S'_{1}(0)+900S'_{3}-784S'_{5}$, leading to 
\begin{align}
	f'
    =\frac{\epsilon}{\Ms^2}
	\bigl\langle\,
			&	y\left(
					1176 y^5-980 y^4-900 y^3+675 y^2+102 y-51
				\right)
    \bigr\rangle_\uv
    \no\\
	&-
    \bigl\langle 
                4y(196 y^4-225 y^2+51)\J^2
	\bigr\rangle_\uv
	\,.\label{f'FESRmod}
\end{align}
Then, we can use \eqref{f'FESRmod} instead of \eqref{alphaSR2a} to constraint $f'/f$. In this way, we can derive sum rules for $f'$ and $\alpha''$. This analysis suggests the existence of FESRs for Regge parameters even in the presence of sub-leading terms represented by $\scat_\text{sub}$, thanks to the multiple FESRs.  
Although the bound on $f'/f$ obtained from \eqref{f'FESRmod} will be slightly weaker than \eqref{alphaSR2a} (by a factor $\sim 7$), the essential point that $f'/f$ is bounded from above by the scale $\Ms$ remains unchanged. 
We therefore assume in this study that the sub-leading terms can be ignored and the consequences of FESRs such as \eqref{alphaSR2a} are discussed under this assumption. More careful analysis on this point in more generic setup is left for future work.  

\subsection{Inclusion of loops of light particles}\label{sec:loop}
Loop contributions from light particles are not taken into account in the previous sections. It is useful to investigate how we can extend our analysis to include such loop corrections. For this purpose, we discuss an extension of our FESRs to the case where the loops of light particles are included. To avoid the IR divergence issues, we consider the situation where the massive light particles are coupled to the massless scalar field $\phi$ and and discuss their contributions to the $\phi\phi\to\phi\phi$ amplitude $\scat(s,t)$. Once such loop corrections are included, the location of the branch cuts of $\scat$ is given by the mass scale of light massive particles $\mth^2$, rather than the mass scale of the lightest higher-spin state $\Ms^2$. Since the starting point \eqref{FESRorigin} of the derivation of the FESRs remains unchanged, our FESRs in the current setup are simply given by \eqref{FESR1} with replacing $S_{2n+1}(t)$ by $\tilde S_{2n+1}(t)$ as
\begin{align}
	\tilde S_{2n+1}(t)
	=
	\frac{f(t)}{\alpha(t)+2n+2}
	\qquad 
	(n=0,1,2,\cdots)
	\,,\label{FESR1loop}
\end{align}
where $\tilde S_{2n+1}(t)$ is defined by  
\begin{align}
	&\tilde S_{2n+1}(t)
	\coloneqq 
		\left[
			\Ms^2/\epsilon+(t/2)
		\right]^{-2n-2}
	\int^{\Ms^2/\epsilon}_{\mth^2}\,\mathrm{d}s\,
		\left(
			s+(t/2)
		\right)^{2n+1} 
	\im\,\scat(s,t)
	+ p_n(t) 
	\,.\label{eq:Sdefloop}
\end{align}
Only the difference between the definition \eqref{eq:Sdef} of $S_{2n+1}(t)$ and that of $\tilde S_{2n+1}(t)$ is the value of the lower end of the integral. Consequently, we can extend other FESRs derived in the previous sections to those in the presence of loops of light particles by simply replacing the value of the lower end of the integral by $\mth^2$. This is equivalent to perform the following replacement in the results obtained in the previous sections:
\begin{align}
    \Ms^2\to \mth^2\,,
    \qquad
    \epsilon \to \delta 
    \coloneqq (\mth^2/\Ms^2)\,\epsilon
    \,.\label{replace}
\end{align}
For instance, we can derive the FESR for $f'$ as 
\begin{align}
	&f'
	= \frac{\delta}{\mth^2}
	\left\langle\,
				y\left(
					-36y^3+27y^2+8y-4
				\right)
			+y(18y^2-8)\J^2
	\right\rangle_\uvir
	\,,\label{alphaSR2aloop}
\end{align}
where 
\begin{align}
	\llangle(\cdots)\rrangle_\uvir
	&\coloneqq 
	\int^{\mth^2/\delta}_{\mth^2}\,\frac{\mathrm{d}s}{s}
	\sum_{\text{even}J\geq0}n_J\rho_J(s) \,(\cdots)
	\label{iruvavs}\\
	&=
	\int^{1}_{\delta}\,\frac{\mathrm{d}y}{y}
	\sum_{\text{even}J\geq0}n_J\rho_J(y\Ms^2/\epsilon) \,(\cdots)
	\,,\quad y= \frac{s}{\mth^2/\delta}
	\,.\label{iruvavy}
\end{align}
Eq.~\eqref{alphaSR2aloop} reduces to \eqref{alphaSR2a} when ignoring loop contributions from light particles. Similarly, we can also derive FESRs for $f$, $\alpha'$, and $\alpha''$ in the current setup. 
In this way, we can easily generalize the FESRs obtained in the previous sections to include loop corrections from light particles.

In the previous section, we derived bounds on the Regge parameters $f'$ and $\alpha''$ with ignoring the loop contributions, and eventually showed \eqref{eq:scaling1}. Its naive generalization to loops would imply
\begin{align}
    M\gtrsim\min\left[\mth, \mth^2\sqrt{\alpha'} \right]\,, \label{scalingloop}
\end{align}
but this does not mean that $M$ has to always be the IR scale $\mth$ since our argument just provides a lower bound on the energy scale $M$. It would be interesting to improve our FESR analysis taking into account more detailed information of the IR physics. We leave this direction for future work.

\subsection{Summary of the working assumptions}\label{sec:assumptions}
It is useful to summarize the assumptions we imposed in this study. The following general properties (i)-(iii) and one techinical property (iv) are assumed  
up to $\mathcal{O}(\Mpl^{-2})$:~\footnote{As a technical remark, we stress that the assumption made in the second line of (i) may be eliminated without changing our main conclusion. This is motivated by the discussions in sec.~\ref{sec:subleading}. It is essential here that we have {\it multiple} FESRs.} 
\begin{itemize}
    \item [(i)] {\it Regge behavior} of $\scat(s,t)$ at $s\geq \Mstar^2\gg\Ms^2$ due to a tower of higher-spin states,  
    and the dominance of the Regge pole term with the spin-2 Regge intercept at $s=\Mstar^2$. 
    \item [(ii)] {\it Unitarity and analyticity} of $\scat(s,t)$ except for usual poles and cuts, which imply the validity of $\rho_J(s)\geq 0$ and the FESRs combined with (i).
      \item [(iii)] {\it Crossing symmetry} of $\scat(s,t)$, implying the $s\text{-}t\text{-}u$ permutation invariance of $\scat$.
    \item [(iv)] {\it Negativity of IR part of null constraints}, {\it i.e.,} the validity of \eqref{nullapp} for $\epsilon=\Ms^2/\Mstar^2\ll1$.
\end{itemize}
The property (i) is the assumption that the behavior of amplitude $\scat(s,t)$ at high energies above $s=\Mstar^2$ is softened by the Reggeization of graviton exchange due to the tower of higher-spin states. 
This property is a natural consequence of the dispersion relation (see around eq.~\eqref{eq:disp1}) and indeed satisfied in the perturbative string theory. The property (iii) is used for deriving null constraints and also for just simplifying the sum rules. The property (iv) is postulated to implement the null constraints into our FESRs. This is suggested by the crossing symmetry and some simple physical intuition as discussed in sec.~\ref{sec:null}. We confirm that the property (iv) is satisfied in known examples in app.~\ref{sec:checkappnull}.

The properties (i) and (ii) allow us to formulate the FESRs for the Regge parameters and the gravitatinoal positivity bound \eqref{gravposiFESR1} only in terms of the physics below the Reggeization scale $\Mstar^2$.
In particular, if the hierarchy between $\Ms$ and $\Mpl$ exists and gravity is UV completed by the Reggeization within its weakly-coupled regime, it is not necessary to concern about the graviton loops. We can also consider a scenario in which the hierarchy between $\Ms$ and $\Mpl$ is absent but strongly-coupled UV completion of gravity is achieved by the Reggeization near the Planck scale. Our formulation may apply to this case too although in this case, it will be necessary to take into account the graviton loop corrections and the associated IR divergent issues should be correctly treated. Graviton loops also modify the analytic structure of $\scat$. 
It would be interesting to consider such scenario.

\section{Conclusion}\label{sec:concl}
We considered two-to-two scattering of an identical massless scalar coupled to gravity up to $\mathcal{O}(\Mpl^{-2})$ as the simplest example of matter-matter scatterings in the presence of gravity. 
We developed a method to constrain the Regge amplitude of graviton exchange by means of the {\it multiple} finite energy sum rules (FESRs) given in eqs.~\eqref{FESR1}, which directly connect gravitational Regge amplitudes at a finite ultraviolet scale with infrared physics.
Although the $s\text{-}t\text{-}u$ permutation invariant process is considered in the present analysis, it is straightforward to derive FESRs even for non crossing-symmetric processes as mentioned in sec.~\ref{sec:FESRderiv}. 
It is demonstrated explicitly that our FESRs are satisfied by string amplitudes. The assumptions we imposed in the present analysis are summarized in sec.~\ref{sec:assumptions}.

The FESRs relate the Regge parameters to the dispersive integral in the low-energy regions below the energy scales of the Reggeization. We used the null constraints which follow from the crossing symmetry as an input data of particle spectrum at such low-energy regimes. We then derived constraints on the Regge parameters, particularly $f'$ and $\alpha''$, leading to the IR finite gravitational positivity bounds on the coefficient $c_2(0)$ of the $s^2$ term in the IR amplitude (see eq.~\eqref{IRexp} for its definition) via \eqref{eq:disp2} in four spacetime dimensions. Our result is complementary to the bounds derived by using the novel finite impact parameter-space method~\cite{Caron-Huot:2021rmr,Caron-Huot:2022ugt}, which suffer from the logarithmic dependence on the IR cutoff in four spacetime dimensions. 

Concretely, our bound on $c_2(0)$ is \eqref{F0bound} with \eqref{Akapp} and \eqref{Bkapp} when the loops of light particles are ignored. We also derived bounds in higher dimensions $D>4$: see {\it e.g.}, \eqref{F0bound_d} and \eqref{akbkvalue_d} for final results. Our bounds involve a positive parameter $\epsilon=\Ms^2/\Mstar^2\ll1$. For the discussions of $\epsilon$-dependence/independence of the bound on $c_2(0)$, see sec.~\ref{sec:review}. When we take $\epsilon=0.01$ as a benchmark point, the bound becomes \eqref{F0boundnum}. Since the $\epsilon$-dependence is weak, we conclude that the relation \eqref{eq:scaling1} is shown, {\it i.e.,} the unknown scale $M$ which parameterize $f'/f$ and $\alpha''/\alpha'$, is bounded by $\Ms$, the mass of lightest higher-spin states. 

We also discussed the case where loops of light particles are included in sec.~\ref{sec:loop}. We derived FESRs in the presence of loops of light particles. By simply extending the analysis in sec.~\ref{sec:FESRbound} to the loop level, we found that our bound on the scale $M$ is given by the weaker one \eqref{scalingloop} instead of \eqref{eq:scaling1}. This however does not necessarily mean that the scale $M$ has to always be identical to the mass scale $\mth$ of the light particles, simply because it just provides a lower bound on the energy scale $M$.
Given its phenomenological relevance (see {\it e.g.,} \cite{Cheung:2014ega,Andriolo:2018lvp,Chen:2019qvr,Alberte:2020jsk,Alberte:2020bdz,Aoki:2021ckh,Noumi:2021uuv,Noumi:2022ybv,Noumi:2022zht}), it is important to develop a method to identify $M$ for a given scattering process and a UV completion scenario, e.g., by improving our FESR analysis taking into account such detailed information. Also, it is in principle straightforward to extend our FESRs to generic scattering processes. From these perspectives, our formulation provides a basic framework for the further study of gravitational positivity bounds in four spacetime dimensions.

\bigskip
\noindent
{\bf Note added:} After we completed the project, Ref.~\cite{deRham:2022gfe} appeared on arXiv, which also used the finite energy sum rule to give constraints on gravitational Regge amplitudes.

\acknowledgments
We thank Yu-tin Huang for useful discussion.
T.N. was supported in part by JSPS KAKENHI Grant No.~20H01902 and No.~22H01220, and MEXT KAKENHI Grant No.~21H00075, No.~21H05184 and No.~21H05462.
J.T. was supported by IBS under the project code, IBS-R018-D1, and JSPS KAKENHI Grant No.~20J00912 and No.~21K13922.

\appendix
\section{Concrete expressions of null constraints: examples}\label{sec:nullexp}
Here, we show the explicit expressions of  $\nullc_k(\J^2)$ with $k=3,4,5,6$ as examples. 
\begin{align}
	&\nullc_3=({(\J^2)}-6) {(\J^2)} (2 {(\J^2)}-49)\,, \\
	&\nullc_4=2 ({(\J^2)}-38) ({(\J^2)}-16) ({(\J^2)}-6){(\J^2)} \,,\\
	&\nullc_5=({(\J^2)}-6) {(\J^2)} \left(2 {(\J^2)}^3-193 {(\J^2)}^2+5958 {(\J^2)}-61560\right)\,, \\
	&\nullc_6=2 ({(\J^2)}-6) {(\J^2)} \left({(\J^2)}^4-154 {(\J^2)}^3+8434 {(\J^2)}^2-194460 {(\J^2)}+1522800\right)\,.
\end{align}

\section{Illustrative examples for IR part of null constraints}\label{sec:checkappnull}
We investigate the validity of \eqref{nullapp} in concrete examples. Only the case $k=3$ is discussed just for simplicity. We consider the type II closed superstring amplitude and the scalar box amplitude in sec.~\ref{sec:checkappnullvs} and \ref{sec:checkappnullbox}, respectively. 
\subsection{Type II superstring amplitude}\label{sec:checkappnullvs}

\begin{figure}[tbp]
 \centering
  \includegraphics[width=.75\textwidth, trim=100 180 170 140,clip]{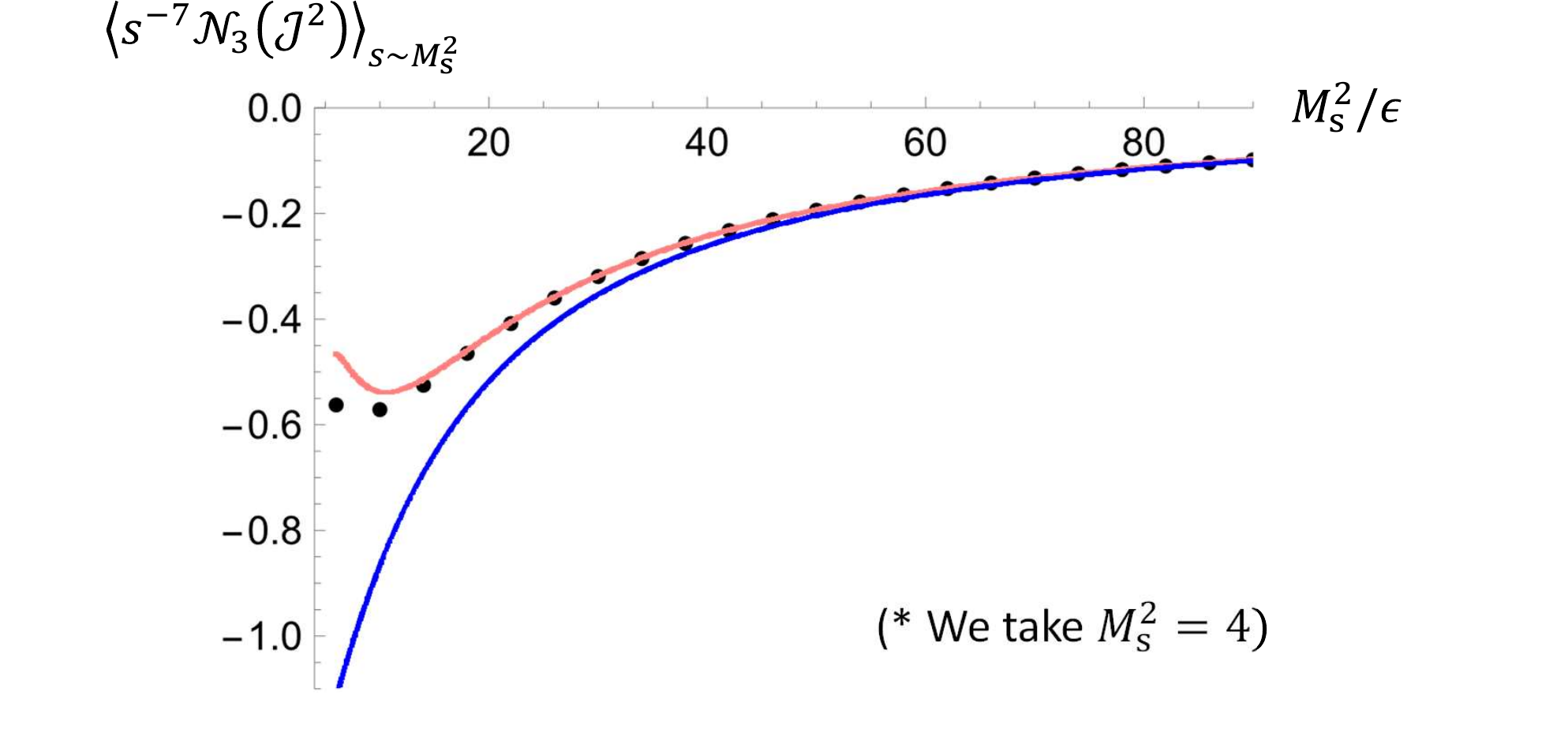}
 \caption{The LHS of \eqref{nullapp} for the type II superstring amplitude. The numerical results obtained by using \eqref{imvs} are expressed by the black dots. The colored lines show the numerical evaluations based on the Regge behavior. The difference between the pink line and the blue line is that the blue line is obtained when the subleading terms are ignored by setting $c_{3,2}=c_{3,1}=0$. To obtain the black dots, we perform the computation for given $\epsilon$ and our choice of $\epsilon$ is $(1/\epsilon)=q+(1/2)$ with $q\in\mathbb N$, following the mid-point prescription of \cite{Ademollo:1968cno,Veneziano:1968yb}.  The plots show that \eqref{nullapp} is indeed satisfied.  The solid lines converge to the series of data points for sufficiently tiny $\epsilon(\lesssim0.1)$, implying the validity of the approximation \eqref{reggevs} and the dominance of the $n=k(=3)$ term in the RHS of \eqref{uvint} for small $\epsilon$.}
 \label{fig:IRnullVS} 
\end{figure}

From the four-dimensional perspective, the ten-dimensional graviton polarized in the extra dimensions can be regarded as a massless scalar. We consider four-point scattering of such a four-dimensional massless scalar in type II closed superstring theory at the tree-level:
\begin{align}
    \scat_\text{type II}(s,t)
    &=
    -\frac{1}{\pi}(s^2u^2+t^2u^2+s^2t^2)\,
    \frac{\Gamma(-s/4)\Gamma(-t/4)\Gamma(-u/4)}{\Gamma(1+s/4)\Gamma(1+t/4)\Gamma(1+u/4)}
    \,,\label{vs}
\end{align}
which we call the type II superstring amplitude in short. Here an irrelevant proportionality constant is omitted and we take $\Ms^2=4$. In the regime $s>0$ and $t\sim0$, this amplitude has only $s$-channel poles,
\begin{align}
    \im\,\scat_\text{type II}(s,t)|_{s>0,\,t\sim0}
    =
    4(s^2u^2+t^2u^2+s^2t^2)
    \sum_{j=0}^\infty
    \left(
        \frac{\Gamma(j+t/4)}{\Gamma(1+t/4)\Gamma(1+j)}
    \right)^2
    \delta(s-4j)
    \,.\label{imvs}
\end{align}
The high-energy behavior of $\im\,\scat_\text{R}$ can be captured by the Regge amplitude. We perform the smearing procedure by considering the limit $|s|\to\infty$ with $0<\text{arg}(s)\ll1$ and fixed $t$ in \eqref{vs} to observe the Regge behavior of the form,
\begin{align}
    \im\,\scat_\text{type II}(s,t)|_{s\gg 4,\, t\sim0}
    \simeq
    \frac{256}{[\Gamma(1+t/4)]^2}\left(\frac{s+t/2}{4}\right)^{2+t/2}
    \,.\label{reggevs}
\end{align}
We evaluate the LHS of \eqref{nullapp} for given $\epsilon$ with $k=3$ by using the exact form \eqref{imvs} or the Regge amptliude \eqref{reggevs}. In the latter case, we use the relation, 
\begin{align}
    \langle s^{-4-k}\nullc_k\rangle_\uv
    =
    - \int^\infty_{\Ms^2/\epsilon}\mathrm{d}s\,s^{-5-k}\sum_{n=1}^kc_{k,n}(s\der_t)^n\im\,\scat(s,t)|_{t=0}
    \,,    \label{uvint}
\end{align}
and substitute \eqref{reggevs} into the integral on the RHS because the Regge behavior is a good approximation only at high energies. 
The results are shown in fig.~\ref{fig:IRnullVS},
which shows that \eqref{nullapp} is indeed satisfied. The validity of the approximation \eqref{reggevs} and the dominance of the $n=k(=3)$ term in the RHS of \eqref{uvint} for small $\epsilon$ are also demonstrated.

 \subsection{Scalar box diagram}\label{sec:checkappnullbox}
\begin{figure}[tbp]
 \centering
  \includegraphics[width=1\textwidth, trim=30 170 20 140,clip]{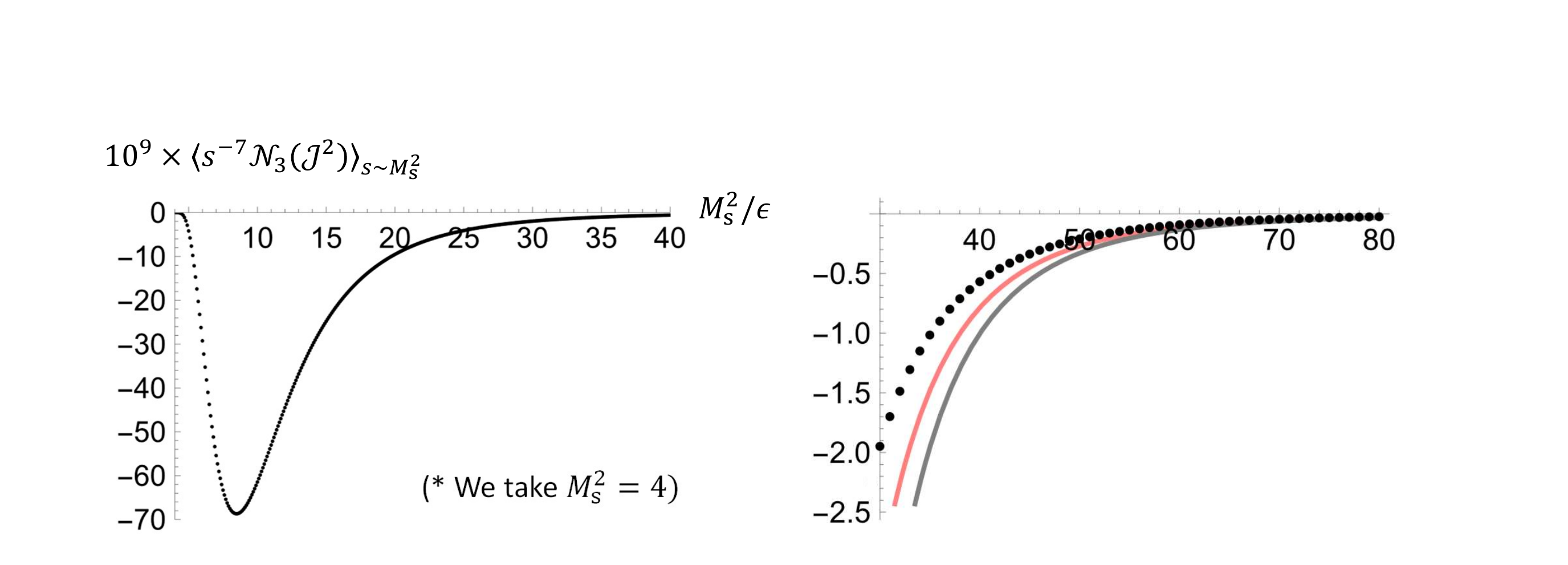}
  \caption{The LHS of \eqref{nullapp} multiplied by the factor $10^9$ for the scalar box amplitude. Numerical results based on the exact form of $\scat_\text{box}$ are shown as the black dots in the both of the figures. To draw these plots, We set $(4/\epsilon)=4,4.1,4.2,\cdots,40$ in the left figure and $(4/\epsilon)=30,31,\cdots,80$ in the right figure. The solid lines on the right figure are obtained by using the high-energy approximation \eqref{boxhigh}. A difference between these lines is that we ignore the subleading terms by setting $c_{3,1}=c_{3,2}=0$ when drawing a gray line. The solid lines converge to the numerical plots in a good approximation for sufficiently small $\epsilon$, implying the validity of the high-energy approximation \eqref{boxhigh} and the dominance of $n=k(=3)$ in the RHS of \eqref{uvint} for small $\epsilon$.}
 \label{fig:IRnullbox} 
\end{figure}

Next, let us demonstrate how the relation \eqref{nullapp} holds in the context without gravity. As an illustrative example, we consider the scalar box amplitude. In particular, we consider massless external line and massive internal line with mass $m$. Omitting irrelevant proportionality constants, the amplitude can be evaluated as
\begin{align}
    \im\,\scat_\text{box}|_{s\gg m^2,\,t\sim0}
    \propto
    \frac{30m^4+5m^2t+t^2+\cdots}{30m^2s}
    +\mathcal{O}(s^{-2})
    \,.\label{boxhigh}
\end{align}
This shows $s^{-5-k}(s\der_t)^n\im\,\scat_\text{box}\sim s^{-6-(k-n)}$ at $s\gg m^2$. We then expect the validity of \eqref{nullapp}. We write $\Ms^2=4m^2$ for simplicity, where we just denote the location of the normal threshold by $\Ms^2$. We take $\Ms^2=4$ and numerically compute the LHS of \eqref{nullapp} for given $\epsilon$ by using the exact form of $\im\,\scat_\text{box}$: the results are shown in the left panel of fig.~\ref{fig:IRnullbox}. This shows that \eqref{nullapp} is indeed satisfied by the scalar box amplitude. 

We also compute the LHS of \eqref{nullapp} by using the high-energy approximation \eqref{boxhigh}: the results are shown in the right panel of fig.~\ref{fig:IRnullbox}. As comparison, the numerical results based on the exact form of $\im\,\scat_\text{box}$ are also shown. The results again confirm that \eqref{nullapp} is indeed satisfied. Also, the validity of the high-energy approximation \eqref{boxhigh} and the dominance of $n=k(=3)$ in the RHS of \eqref{uvint} for small $\epsilon$ are shown.

\section{FESRs for the type II superstring amplitude} \label{sec:FESRconsitency}
We demonstrate how the type II superstring amplitude~\eqref{vs} satisfies FESRs for $f$, $f'$, $\alpha'$, and $\alpha''$. For this purpose, by comparing \eqref{reggevs} with the parameterization \eqref{eq:regge1}, we obtain $f(t)$ and $\alpha(t)$ of the type II superstring amplitude as
\begin{align}
    f(t) = \frac{256}{[\Gamma(1+t/4)]^2}\left(\frac{1}{\epsilon}+\frac{t}{8}\right)^{\alpha(t)}
    \,,\qquad
    \alpha(t) = 2 + t/2
    \,.\label{VSregge}
\end{align}
In particular, we have 
\begin{align}
    f\epsilon^2=256
    \,,\quad
    f'/f = \frac{1}{4}
    \left[
        2\gamma+2\ln(1/\epsilon)+\epsilon 
    \right] 
    \,,\qquad
    \alpha'= \frac{1}{2}
    \,,\quad
    \alpha''=0
    \,. \label{VSreggeparameter}
\end{align}
Here, $\gamma\approx 0.5772$ is the Euler–Mascheroni constant. 
Our interest in this section is if these values are reproduced via FESRs such as \eqref{alphaSR2a} and \eqref{f'FESR2} by evaluating the RHS of the equations for given $\epsilon$. We parameterize $\epsilon$ as $(1/\epsilon)=q+(1/2)$ with $q\in\mathbb N$, as we did in sec.~\ref{sec:checkappnull}. To clearly show the validity of FESRs, we choose $q=1,2,\cdots 14$ in the analysis of $f$ while we choose $q=1,3,5,\cdots 29$ in the analysis of $f'/f$ and $\alpha'$. We choose $q=1,3,5,\cdots 59$ for $\alpha''$. 
We confirm below that the predictions made by FESRs converge to the correct values of Regge parameters \eqref{VSreggeparameter} for tiny $\epsilon$ as expected. 

\begin{figure}[tbp]
 \centering
  \includegraphics[width=.9\textwidth, trim=50 160 70 140,clip]{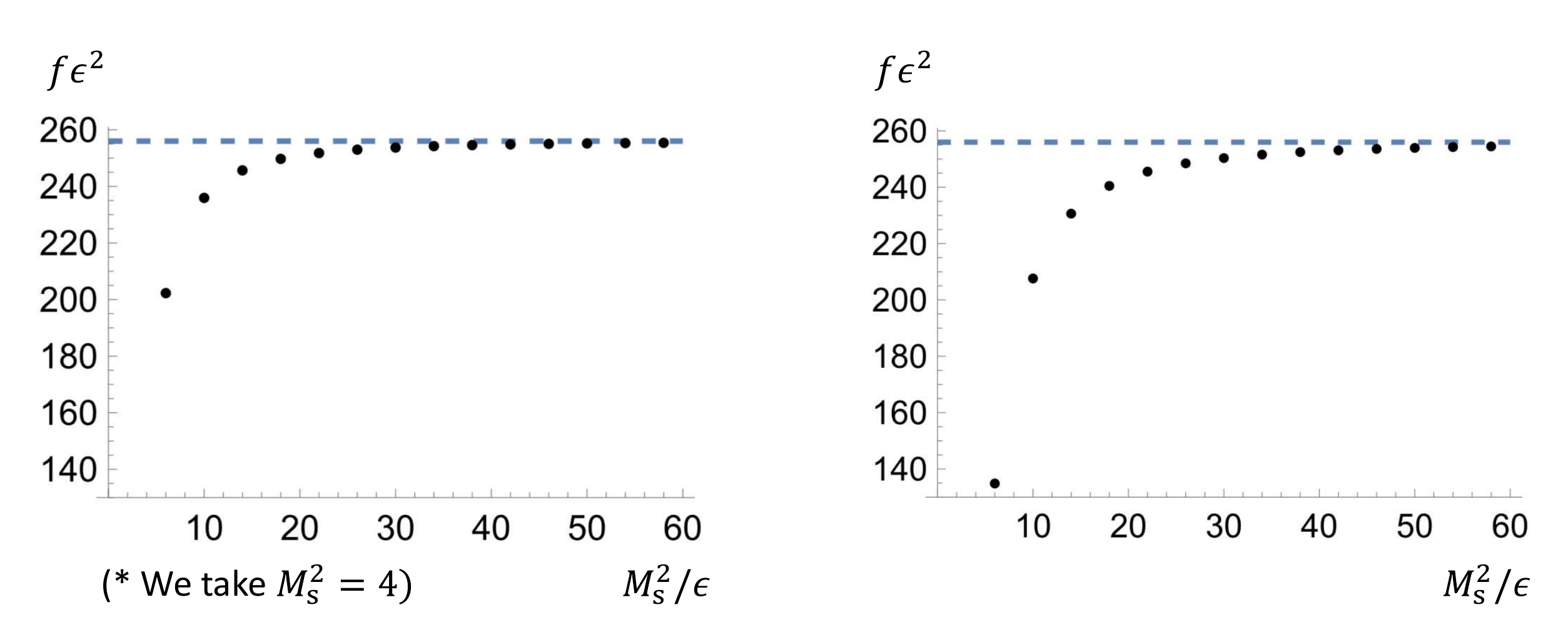}
 \caption{Predictions of the FESRs \eqref{FESRfwd1} with $n=0$ and $n=1$ for $f\epsilon^2$ of $\scat_\text{type II}$ are plotted as black dots on the left and the right panel, respectively. A choice of $\epsilon$ is $(1/\epsilon)=q+(1/2)$ with $q=1,2,\cdots, 14$. A correct value of $f\epsilon^2$ is drawn as a blue dashed line. The FESR predictions match well with the correct value for sufficiently tiny $\epsilon$ as expected.}
 \label{fig:fFESRconsistency} 
\end{figure}
\begin{figure}[tbp]
 \centering
  \includegraphics[width=.95\textwidth, trim=50 175 20 120,clip]{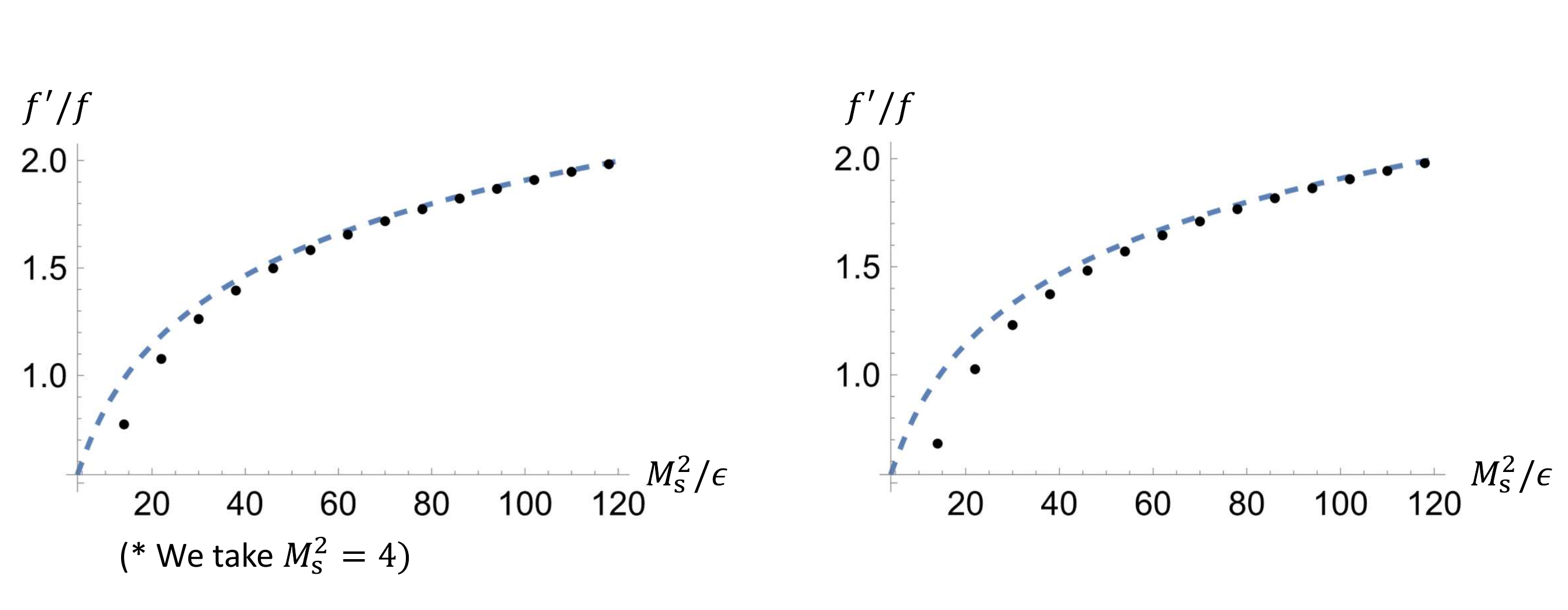}
 \caption{Predictions of the FESRs \eqref{alphaSR2a} and \eqref{f'FESR2} for $f'/f$ of $\scat_\text{type II}$ are plotted as black dots on the left and the right panel, respectively. A choice of $\epsilon$ is $(1/\epsilon)=q+(1/2)$ with $q=1,3,5,\cdots, 29$. A correct value of $f'/f$ is drawn as a blue dashed line. This figure again confirms that FESR predictions match well with the correct value when taking $\epsilon$ to be sufficiently as expected.}
 \label{fig:f'FESRconsistency} 
\end{figure}

\begin{figure}[tbp]
 \centering
  \includegraphics[width=.8\textwidth, trim=20 210 25 160,clip]{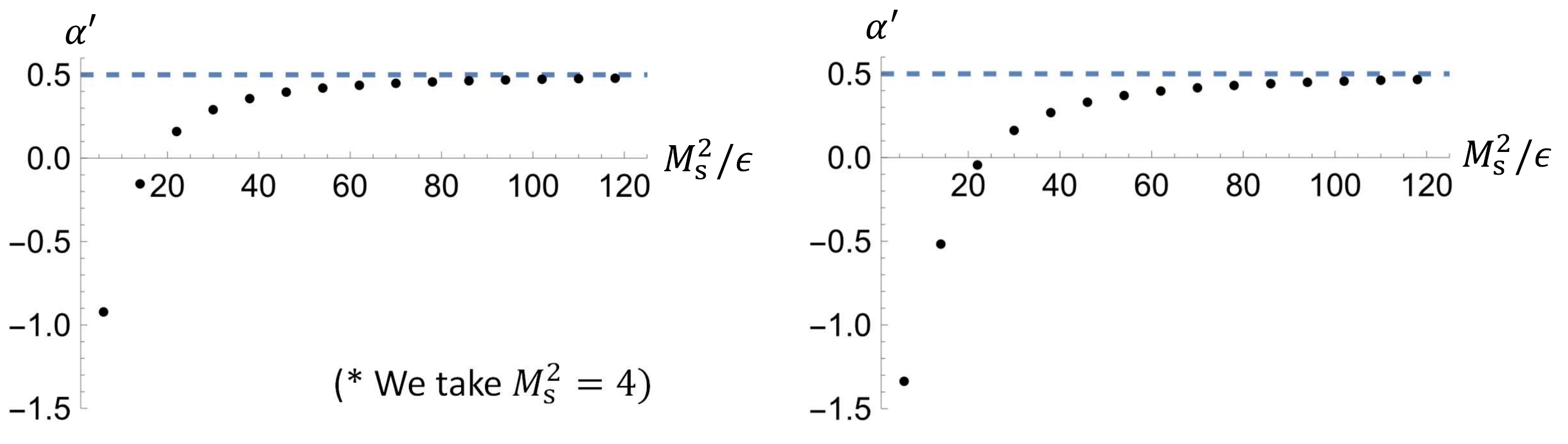}
 \caption{Predictions of the FESR \eqref{ddotalphabd1} for $\alpha'$ of $\scat_\text{type II}$ are plotted as black dots. A choice of $\epsilon$ is $(1/\epsilon)=q+(1/2)$ with $q=1,3,5,\cdots,29$. A correct value is $\alpha'=1/2$, drawn as a blue dashed line. The FESR predictions converge to the correct value for sufficiently tiny $\epsilon$ as expected.}
 \label{fig:alpha'FESRconsistency} 
\end{figure}

\begin{figure}[tbp]
 \centering
  \includegraphics[width=.6\textwidth, trim=100 140 200 210,clip]{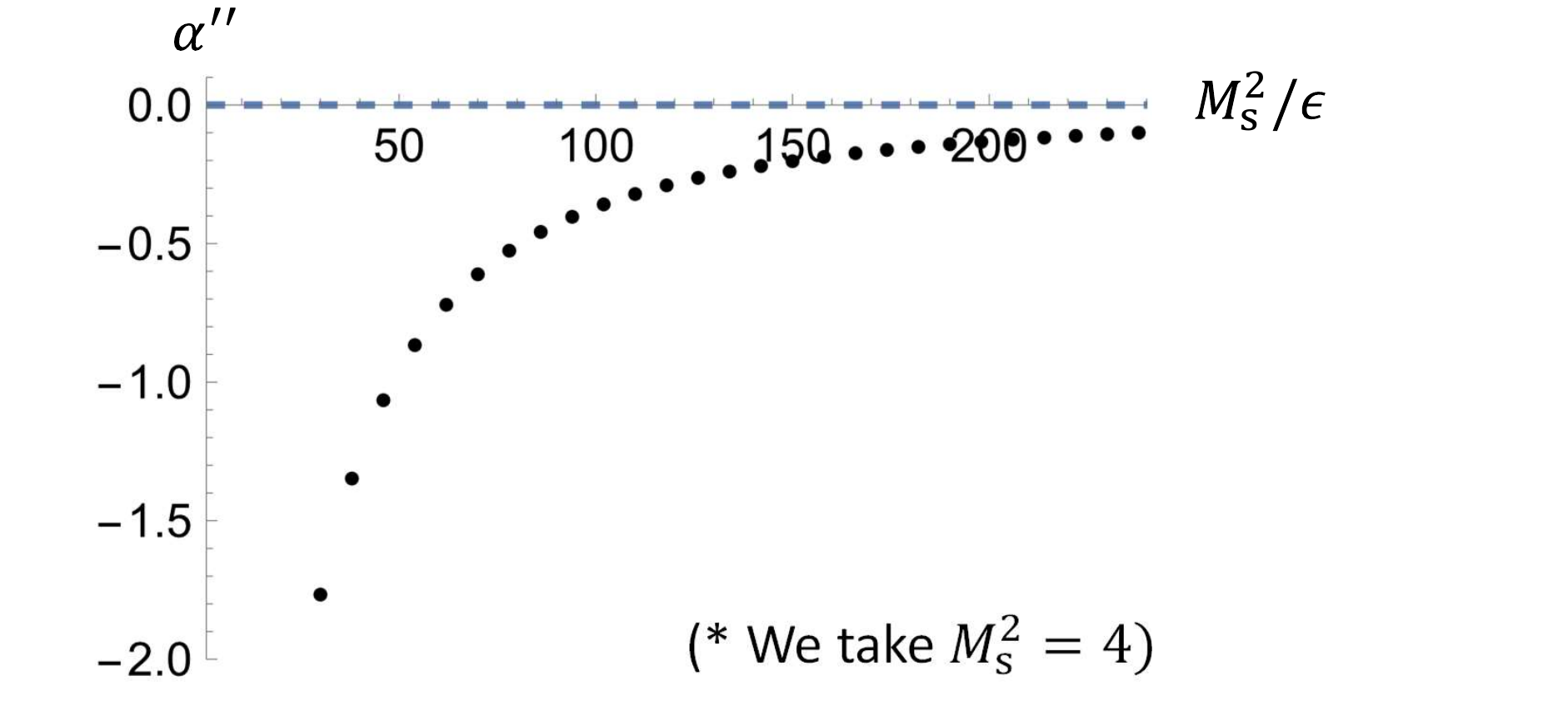}
 \caption{Predictions of the FESR \eqref{ddotalphabd1} for $\alpha''$ of $\scat_\text{type II}$ are plotted as black dots. A choice of $\epsilon$ is $(1/\epsilon)=q+(1/2)$ with $q=1,3,5,\cdots,59$. A correct value is $\alpha''=0$, drawn as a blue dashed line. The FESR predictions converge to the correct value for sufficiently tiny $\epsilon$ as expected.}
 \label{fig:alpha''FESRconsistency} 
\end{figure}

\paragraph{Reproduction of $f$ and $f'/f$.} We use the FESRs for $f$ , particularly \eqref{FESRfwd1} with $n=0$ and $n=1$ cases, to predict the value of $f$. The results for given $\epsilon$ are plotted in fig.~\ref{fig:fFESRconsistency}.  
Similarly, we also compute the predictions of FESRs \eqref{alphaSR2a} and \eqref{f'FESR2} for $f'/f$: the results are shown in fig.~\ref{fig:f'FESRconsistency}. In both cases, we find a good agreement with the correct values when taking $\epsilon$ to be sufficiently small.

\paragraph{Reproduction of $\alpha'$ and $\alpha''$.} 
We use \eqref{alphaSR1b} with $(m,n)=(1,0)$ and $(m,n)=(2,0)$, the FESRs for $\alpha'$, to predict the value of $\alpha'$. The results are shown in fig.~\ref{fig:alpha'FESRconsistency}. Similarly, we also evaluate the prediction of FESR \eqref{ddotalphabd1} for $\alpha''$: the results are shown in fig.~\ref{fig:alpha''FESRconsistency}. In both cases, we find a good agreement with the correct values when taking $\epsilon$ to be sufficiently small.

Also, the results shown in figs.~\ref{fig:fFESRconsistency}, \ref{fig:f'FESRconsistency}, and \ref{fig:alpha'FESRconsistency} demonstrate that different FESRs for the same quantity such as eqs.~\eqref{alphaSR1a} and \eqref{f'FESR2} are consistent with each other. This implies that the consistency condition \eqref{consistency1} is also satisfied.

\section{Derivation of a lower bound on $\alpha''$}\label{sec:boundalpha''detail}
In this section, we derive eqs.~\eqref{alpha''bound1} and \eqref{Bkapp}.
\subsection{Numerical evaluation}
We first set $n=1$ in \eqref{ddotalphabd3} because this choice gives the best bound. The reason for this is precisely analogous to the one mentioned in sec.~\ref{sec:boundf'}. From now on, we determine $B_{1,k}(\gamma)$ and choose a positive free parameter $\gamma$ appropriately to maximize $B_{1,k}(\gamma)$. 
As in the case of $I_{1,k}(y,\J^2;\beta)$, the behavior of $\mathcal{H}_{1,k}(y,\J^2;\gamma)$ in the small-$J$ region $0\leq J \leq J_k$ differs from the one in the large-$J$ region $J\geq J_k+2$. We discuss these two cases separately. Below, we consider $k=3,4,5,\cdots,12$.

\paragraph{Small $J$ analysis.}
We evaluate the minimum value of $\mathcal{H}_{1,k}(y,\J^2;\gamma)$ within the region $y\in[\epsilon,1]$ and $J\in\{0,2,4,\cdots J_k\}$. 
$\mathcal{H}_{1,k}(y,\J^2;\gamma)$ is dominated by the final term on the RHS of \eqref{Hdef} when $\epsilon$ is sufficiently tiny and $\gamma$ is not so small to satisfy $\gamma\gtrsim\epsilon^{4+k}$. Then, for such $\epsilon$ and $\gamma$, $\mathcal{H}_{1,k}(y,\J^2;\gamma)$ will be minimized at $(y,J)=(\epsilon,J_{*,k})$:
\begin{align}
	B_{1,k}^{\text{small-$J$}}(\gamma)
	\coloneqq
	\mathop{\min_{J\in\{0,2,\cdots,J_k\}}}_{y\in[\epsilon,1]}\mathcal{H}_{1,k}(y,\J^2;\gamma)
	=
	\mathcal{H}_{1,k}(\epsilon,\J^2_{*,k};\gamma)\,.\label{HlowJmin}
\end{align}
Assuming that \eqref{HlowJmin} is true, we choose  $\gamma$ to optimize the bound. We can then confirm that eq.~\eqref{HlowJmin} is indeed valid for such $\gamma$. $B_{1,k}^{\text{small-$J$}}(\gamma)$ is a  monotonically decreasing function of $\gamma$. 

\paragraph{Large $J$ analysis.}Next, we consider the $J\geq J_k+2$ case. We first use inequalities $a_1(y)> -268y$ and $a_2(y)> -2\left(81-8\sqrt{70}\right)y$ which are valid within the region $y\in[\epsilon,1]$ to get 
\begin{align}
	\mathcal{H}_{1,k}(y,\J^2;\gamma)
	> 
	-2\left(81-8\sqrt{70}\right)(\J^2)^2-154\J^2+\gamma y^{-6-k}\nullc_k(\J^2)
	\,. 
\end{align}
The RHS is minimized at $y=1$ because $\nullc_k(\J^2)>0$ for $J\geq J_k+2$. Also, we treat $J$ as continuous variables for convenience. Then, we have 
\begin{subequations}
    \label{HhighJmin}
\begin{align}
	&\mathop{\min_{\text{even} J\geq J_k+2}}_{y\in[\epsilon,1]}\mathcal{H}_{1,k}(y,\J^2;\gamma)
	> 
	B_{1,k}^{\text{large-$J$}}(\gamma)\,,
        \\
&        B_{1,k}^{\text{large-$J$}}(\gamma)
       \coloneqq
	\min_{J\geq J_k+2}
		\left[
			-2\left(81-8\sqrt{70}\right)(\J^2)^2-154\J^2+\gamma \nullc_k(\J^2)
		\right]
	\,.
\end{align}
\end{subequations}
The function $-2\left(81-8\sqrt{70}\right)(\J^2)^2-154\J^2+\gamma \nullc_k(\J^2)$ is a polynomial of $\J^2$ of degree $k$ and its minimum can be evaluated analytically.  $-2\left(81-8\sqrt{70}\right)(\J^2)^2-154\J^2+\gamma \nullc_k(\J^2)$ is a monotonically increasing function of $\gamma$, as well as $B_{1,k}^{\text{large-$J$}}(\gamma)$. 

\paragraph{Derivation of $B_{1,k}(\gamma)$.}
We can then derive $B_{1,k}(\gamma)$ from eqs.~\eqref{HlowJmin} and \eqref{HhighJmin} via $
B_{1,k}(\gamma)=\min\bigl[B_{1,k}^{\text{small-$J$}}(\gamma),\,B_{1,k}^{\text{large-$J$}}(\gamma)\bigr]$. 
Because $B_{1,k}^{\text{small-$J$}}(\gamma)$ and $B_{1,k}^{\text{large-$J$}}(\gamma)$ are monotonically increasing and decreasing functions of $\gamma$, respectively, there exists a point $\gamma=\gamma_k^\text{exact}$ for which we have 
\begin{align}
	B_{1,k}^{\text{small-$J$}}(\gamma_k^\text{exact})
	= B_{1,k}^{\text{large-$J$}}(\gamma_k^\text{exact})
	\,.\label{gamma_exact}
\end{align}
Consequently, $B_{1,k}(\gamma)$ is maximized at $\gamma=\gamma_k^\text{exact}$. We solve eq.~\eqref{gamma_exact} numerically for given $\epsilon$. Writing the numerical solution for given $\epsilon$ as $\gamma_k^\text{num}$, we compute $B_{1,k}(\gamma_k^\text{num})$. As examples, we do the analysis for $\epsilon=0.01,0.02,\cdots, 0.3$ with $k=3,4,\cdots,12$ and we check numerically that eq.~\eqref{HlowJmin} is indeed valid for our choice of $(\epsilon,\gamma_k^\text{num})$.

As in the case for the evaluation of $\mathcal A_k$, the numerical computation becomes heavier for higher-$k$. Hence, for $k>12$ cases, we do not perform the numerical evaluation of $\mathcal{B}_k$ and instead we use the analytic expression derived in app.~\ref{sec:analyicalpha}. 

\subsection{Analytic approximation} \label{sec:analyicalpha}
Now we derive an analytic expression of $\mathcal{B}_k$ which is valid at the leading order in small $\epsilon$-expansions. For this purpose, we derive an analytic expression of $B_{1,k}^\text{large-$J$}(\gamma)$ which is valid at the leading order in small $\epsilon$-expansions. When $\gamma$ is sufficiently small, the minimum of $-2\left(81-8\sqrt{70}\right)(\J^2)^2-154\J^2+\gamma \nullc_k(\J^2)$ is realized at $J\gg1$. In this case, we can approximate the function around its minimum as 
\begin{align}
	-2\left(81-8\sqrt{70}\right)(\J^2)^2-154\J^2+\gamma \nullc_k(\J^2)
	\simeq
	-2\left(81-8\sqrt{70}\right)(\J^2)^2+2\gamma(\J^2)^k
	\,,
\end{align}
where the RHS is minimized at $\J^2=\left(\gamma k/2\left(81-8\sqrt{70}\right)\right)^{\frac{1}{2-k}}\gg1$ for sufficiently small $\gamma$. We then obtain a simple but approximate expression of $B_{1,k}^\text{large-$J$}(\gamma)$ as
\begin{align}
	B_{1,k}^\text{large-$J$}(\gamma)
	\simeq
	\frac{-2\left(81-8\sqrt{70}\right)(k-2)}{k}
		\left(
			\frac{2\left(81-8\sqrt{70}\right)}{\gamma k}
		\right)^{\frac{2}{k-2}}
	\,.\label{HhighJminapp}
\end{align}
We can also approximate $B_{1,k}^\text{small-$J$}$ as 
\begin{align}
	B_{1,k}^\text{small-$J$}
	\simeq 
	\gamma\epsilon^{-6-k} \nullc_k(\J^2_{*,k})
	\,.\label{hlowJminapp}
\end{align}
We approximately solve $B_{1,k}^\text{small-$J$}(\gamma)=B_{1,k}^\text{large-$J$}(\gamma)$ by using \eqref{HhighJminapp} and \eqref{hlowJminapp} to get an approximate expression of $\gamma_k^\text{exact}$: the result is 
\begin{align}
	\gamma_k^\text{exact}
	\simeq
	\gamma_k^\text{approx}
	\coloneqq
	\frac{2\left(81-8\sqrt{70}\right)}{k}
		\left(
			\frac{-\nullc_k(\J^2_{*,k})}{k-2}
		\right)^{-1+\frac{2}{k}}
	\epsilon^{4+k-\frac{12}{k}}
	\,.\label{gamma_app}
\end{align}
This shows $\gamma_k^\text{approx}\gg\epsilon^{4+k}$ when $\epsilon$ is sufficiently tiny, providing an analytic understanding why eq.~\eqref{HlowJmin} is indeed valid when choosing $\gamma\sim\gamma_k^\text{exact}$. Substituting \eqref{gamma_app} into $B_{1,k}^\text{small-$J$}(\gamma)$ or $B_{1,k}^\text{large-$J$}(\gamma)$, we obtain an approximate expression of $B_{1,k}(\gamma)$. Finally, we get \eqref{Bkapp}, an approximate expression of $\mathcal{B}_k$.

\section{Extention to general spacetime dimensions}\label{sec:general_d}
In this section, we generalize our analysis to general spacetime dimensions $D\geq4$. We briefly explain how the important relations which are used to derive our bounds in the main text can be generalized to $D$-dimensions. We then derive our bounds \eqref{akbkvalue_d} and \eqref{fpbound}. 

As we will explain just below, the appropriate basis of partial wave expansions depends on $D$. However, this dependence does not change the essential properties of null constraints and FESRs while their precise expressions depend on $D$. Therefore, the logic of our discussions in the main text holds even in higher dimensions $D>4$.

\paragraph{Partial wave expansion.}
The appropriate basis of partial wave expansion depends on $D$: it is given by the Gengenbauer polynomial $P_J^{(D)}$ defined in terms of the Hypergeometric funtion ${}_2F_1$,
\begin{align}
    P_J^{(D)}(z)={}_2F_1\left(-J,J+D-3,\frac{D-2}{2},\frac{1-z}{2}\right)
    \,.
\end{align}
For $D=4$, this reduces to the Legendre polynomial $P_J(z)$. In terms of this basis, the partial wave expansion in $D$-dimensions is written as
\begin{align}
    \im\,\scat(s,t)
	=
	\sum_{\text{even}J\geq0}
	n_J^{(D)}\rho_J^{(D)}(s)P_J^{(D)}
	\left(
		1+\frac{2t}{s}
	\right)
	\,,\quad
	 n_J^{(D)}
	 = \frac{(4\pi)^{\frac{D}{2}}(D+2J-3)\Gamma(D+J-3)}{\pi \Gamma\left(\frac{D-2}{2}\right)\Gamma(J+1)}\,.
\end{align}
Accordingly, we generalize the notation $\langle (\cdots)\rangle_\uv$ given in \eqref{midavs} and \eqref{midavy} as
\begin{align}
	\llangle(\cdots)\rrangle^{(D)}_\uv
	&\coloneqq 
	\int^{\Ms^2/\epsilon}_{\Ms^2}\,\frac{\mathrm{d}s}{s}
	\sum_{\text{even}J\geq0}n_J^{(D)}\rho_J^{(D)}(s) \,(\cdots)
	\label{midavs_d}\\
	&=
	\int^{1}_{\epsilon}\,\frac{\mathrm{d}y}{y}
	\sum_{\text{even}J\geq0}n_J^{(D)}\rho_J^{(D)}(y\Ms^2/\epsilon) \,(\cdots)
	\,,\quad y\coloneqq \frac{s}{\Ms^2/\epsilon}
	\,.\label{midavy_d}
\end{align}
We impose the unitarity condition in the form of $\rho_J^{(D)}\geq0$. 

\paragraph{Null constraints.}
Sum rules \eqref{nullgen1} and \eqref{nullgen_def} are generalized as  
\begin{align}
    &\int^{\infty}_{\Ms^2}\,\frac{\mathrm{d}s}{s}
		\sum_{\text{even}J\geq0}n_J^{(D)}\rho_J^{(D)}(s) \,X_\ell^{(D)}(t;s,J)
	=
	0\qquad
	(\ell=4,6,\cdots)\,,\label{nullgen1_d}
	\\
	&X_\ell^{(D)}(t;s,J)
	\coloneqq 
	\frac{2s+t}{t(s+t)}\frac{P_J^{(D)}(1+\frac{2t}{s})}{(ts(s+t))^{\ell/2}}
	\no\\
	&\qquad\qquad\qquad
	-\mathop{\text{Res}}_{x=0}
		\left[
			\frac{(2s+x)(s-x)(s+2x)}{x(t-x)(s+x)(s-t)(s+t+x)}
			\frac{P_J^{(D)}(1+\frac{2x}{s})}{(xs(x+s))^{\ell/2}}
		\right]
	\,.\label{nullgen_def_d}
\end{align}
As a generalization of $\nullc_k$-constraints in the $D=4$ case, 
we define the null constraints derived from the sum rules $\llangle\der_t^{k-3}X^{(D)}_4(t;s,J)|_{t=0}\rrangle^{(D)}=0$ $(k=3,4,5,\cdots)$ as $\llangle\nullc_{k}^{(D)}(\J^2_D)\,s^{-4-k}\rrangle=0$, where $\J^2_{D}\coloneqq J(J+D-3)$. We impose the normalization condition such that $\nullc_k^{(D)}(\J^2_D)\to2(\J^2_D)^k$ in the large $J$ limit. 

The basic properties of $\nullc_k^{(D)}$ do not depend on $D$: the value of $\nullc_k^{(D)}(\J^2_D)$ can be negative for $J=0,2,4,\cdots, J_k^{(D)}$ while it is always nonnegative for $J\geq J_k^{(D)}+2$. We can check that this is true at least for $D=4,5,\cdots,12$ with $k=3,4,\cdots,24$, but we expect this will be generically correct. We can also check that the value of $J_k^{(D)}$ is independent of $D$ at least for the above mentioned values of $(D,k)$. We introduce an even nonnegative integer $J_{D*,k}$ for which the $\nullc^{(D)}_k(\J^2_D)$ is minimized. Accordingly, we define $\J^2_{D*,k}\coloneqq \J^2|_{J=J_{D*,k}}$. Note that we have $J_{D*,k}=J_k^{(D)}$ at least for the above mentioned values of $(D,k)$.

Because the basic properties of $\nullc_k^{(D)}$ do not depend on $D$, we expect that the discussion of sec.~\ref{sec:appnull} works in general $D\geq4$. Hence, we assume the generalization of \eqref{nullapp}:
\begin{align}
    \langle \nullc_k^{(D)}(\J^2_D)\rangle^{(D)}_\uv
    \leq
    0
    \,.\label{nullapp_d}
\end{align}

\paragraph{FESRs.}
The original FESRs \eqref{FESR1} with the definitions \eqref{eq:Sdef} of $S_{2n+1}(t)$ are valid in general $D$-dimensions. However, if we rewrite them in the form of average $\langle(\cdots)\rangle^{(D)}$, the expression $(\cdots)$ also depends on $D$ in general. This is because the $t$-dependence of $P_J^{(D)}(1+2t/s)$ depends on $D$. For instance, the FESR \eqref{alphaSR2a} for $f'$ takes the following form in $D$-dimensions:
\begin{align}
    &f'
    = \frac{\epsilon}{\Ms^2}
    \left\langle\,
        y\left(
		-36y^3+27y^2+8y-4
	\right)
	+\frac{2}{D-2}\,y(18y^2-8)\J^2_{D}
	\right\rangle^{(D)}_\uv
	\,.\label{alphaSR2a_d}
\end{align}
Similarly, the FESR \eqref{ddotalphabd1} for $\alpha''$ is generalized to
\begin{align}
    &\frac{\alpha''f}{2}
	= -\alpha'f'
	+ 	\frac{\epsilon^2}{\Ms^4}
	\left\langle 
		y\left[a_0(y)+a_1^{(D)}(y)\J_D^2+a_2^{(D)}(y)\J_D^2(\J_D^2-2)\right]
	\right\rangle^{(D)}_\uv
	\,,\label{ddotalphabd1_d}
\end{align}
where
\begin{subequations}
\label{defa_d}
\begin{align}
    &a_1^{(D)}(y)
    =\frac{8 \left(-480 D y^5+80 (3 D+8) y^4+324 D y^3-81 (D+8) y^2-28 D y-14 (D-8)\right)}{(D-2) D y}
    \,,\\
    &a_2^{(D)}(y)
    = \frac{16 \left(80 y^4-81 y^2+14\right)}{(D-2) D y}\,.
\end{align}
\end{subequations}
Note that $a_0(y)$ is given by \eqref{defa0}. By contrast, we can obtain the general expressions of eqs.\eqref{FESRfwd1} and \eqref{originalpartial1} in $D$-dimensions by simply replacing $\langle(\cdots)\rangle$ to $\langle(\cdots)\rangle^{(D)}$:
\begin{subequations}
\label{FESRfwd_d}
\begin{align}
   & S_{2n+1}(0)
    =
    \langle y^{2n+2} \rangle^{(D)}_\uv
    =
    \frac{f}{2n+4} 
    \qquad (n=0,1,2,\cdots)
    \,, \\
    & \frac{f}{4}
    \leq
    \langle y^{1-k} \rangle^{(D)}_\uv
    \leq
    \frac{f}{4\epsilon^{1+k}}
    \qquad
    (k\geq 0)
    \,.
\end{align}
\end{subequations}
This is because we have $P_{J}^{(D)}(1)=1$ for general $D\geq4$.

\paragraph{Bounds on $f'/f$, $\alpha''/\alpha'$, and $c_2$.}
Now we can obtain bounds on $f'/f$, $\alpha''/\alpha'$, and $c_2(0)$ in general $D$ dimensions by following the logical steps explained in sec.~\ref{sec:FESRbound}. We use the FESRs~\eqref{alphaSR2a_d}, \eqref{ddotalphabd1_d}, and \eqref{FESRfwd_d} in addition to \eqref{nullapp_d} and the unitarity condition $\rho_J^{(D)}\geq0$. Let us parameterize bounds on $f'/f$ and $\alpha''/\alpha'$ in general $D$ dimensions as \eqref{reggebound_d}, analogously to \eqref{f'bound1} and \eqref{alpha''bound1}. 
Following the steps explained in sec.~\ref{sec:analyticf}, we obtain an analytic estimate of $\mathcal {A}_k^{(D)}$ as
\begin{align}
	\mathcal{A}_k^{(D)}
	\simeq
	\frac{5(k-1)}{(D-2)k}
		\left(
			\frac{-\nullc_k^{(D)}(\J^2_{D*,k})}{2k-2}
		\right)^{\frac{1}{k}}
	\epsilon^{-\frac{6}{k}}
        \,,\label{Akapp_d}
\end{align}
which is a generalization of \eqref{Akapp}. Let us point out that the prefactor depends on $D$ such that it becomes smaller for larger $D$. This means that our bounds on $f'/f$ are improved for larger $D$ as expected. Technically, the $D$-dependence of the prefactor of \eqref{Akapp_d} comes from the $D$-dependence of the coefficient of the $\J^2_D$ term in \eqref{alphaSR2a_d}.

Similarly, an analytic estimate \eqref{Bkapp} of $\mathcal {B}_k$ is generalized to
\begin{align}
	\mathcal{B}_k^{(D)}
	\simeq
	\frac{4\left(81-8\sqrt{70}\right)}{D(D-2)}\frac{k-2}{k}
		\left(
			\frac{-\nullc_k^{(D)}(\J^2_{D*,k})}{k-2}
		\right)^{\frac{2}{k}}
	\epsilon^{-\frac{12}{k}}
	\,.\label{Bkapp_d}
\end{align}
Again, we find that the prefactor depends on $D$ such that it becomes smaller for larger $D$. This dependence comes from the $D$-dependence of the coefficient of the $(\J^2_D)^2$ term in \eqref{ddotalphabd1_d}. Thanks to this dependence, our bounds on $\alpha''/\alpha'$ are also improved for larger $D$. By using \eqref{Akapp_d} and \eqref{Bkapp_d}, we can obtain the results \eqref{akbkvalue_d} and \eqref{fpbound}.

\bibliography{posi2022.bib}

\end{document}